\documentclass[showpacs,twocolumn,prl,linenumbers]{revtex4}%

\usepackage{amsmath,graphicx}
\usepackage[dvipsnames]{xcolor}\usepackage[colorlinks=true,citecolor=blue]{hyperref}
\usepackage{ulem}

\newcommand{\ket}[1]{\left| #1 \right>} 
\let\baraccent=\= 
\renewcommand{\=}[1]{\stackrel{#1}{=}} 

\newcommand{\Et}{E_{Th}}
\newcommand{\mXp}{$\chi '$}
\newcommand{\Xp}{\chi '}

\newcommand{\X}{\chi }

\newcommand{\mXj}{$\chi_J$}
\newcommand{\Xj}{\chi_J}
\newcommand{\Xd}{\chi_D}

\newcommand{\mXpd}{$\chi'_D$}

\newcommand{\mXpnd}{$\chi'_{ND}$}

\newcommand{\mXs}{$\chi "$}

\newcommand{\Xs}{\chi "}

\newcommand{\tin}{\tau_{in}}

\begin{document}
\title{Coherence-enhanced, phase-dependent   dissipation in long SNS Josephson junctions: revealing Andreev Bound States dynamics 
}

\date{\today}
\author{B. Dassonneville$^{1}$, A. Murani$^{1}$, M. Ferrier$^{1}$,  S. Gu\'eron$^{1}$ and H. Bouchiat$^{1}$ }

\affiliation{$^{1}$ LPS, Univ. Paris-Sud, CNRS, UMR 8502, F-91405 Orsay Cedex, France}

\begin{abstract}

One of the best known causes of dissipation in ac driven quantum systems stems from photon absorption causing transitions between levels. Dissipation can also be caused by the retarded response to the time-dependent excitation, and in general gives insight into the system's relaxation times and mechanisms. 
Here we address the dissipation in a mesoscopic normal wire with superconducting contacts, that sustains a dissipationless supercurrent at zero frequency and that may therefore naively be expected to remain dissipationless at  frequency lower than the superconducting gap. We probe the high frequency  linear response of such a Normal/Superconductor (NS) ring to a time-dependent flux by coupling it  to a highly sensitive multimode microwave resonator. Far from being the simple, dissipationless, derivative of the supercurrent-versus-phase relation, the ring's ac susceptibility also displays  a dissipative component whose phase dependence is a signature of the dynamical processes occurring within the Andreev spectrum. We show how dissipation is driven by the competition between two mechanisms. The first is the relaxation of the  Andreev levels distribution function, while the second correspond to microwave- induced transitions within the spectrum.  Depending on the relative strength of those contributions, dissipation can be maximal at $\pi$ , a phase at which the proximity-induced minigap closes, or can be maximal near $ \pi/2$, a phase at which the dc supercurrent is maximal. We also find that the dissipative response paradoxically increases at low temperature and can even exceed the normal state conductance. The results are successfully confronted with theoretical  predictions  of the Kubo linear response  and time-dependent Usadel equations, derived from the Bogoliubov-de Gennes Hamiltonian  describing  the SNS junction. 
These experiments thus demonstrate the power of the ac susceptibility measurement of individual hybrid mesoscopic systems in probing in a controlled way the quantum dynamics of Andreev bound states. By spanning different physical regimes, our experiments provide a unique access to inelastic scattering and spectroscopy of an isolated quantum coherent system, and reveal the associated relaxation times. This technique should be a tool of choice to investigate topological superconductivity and detect the topological protection of edge   states.
\end{abstract}

\maketitle

\section{Introduction}
 Phase coherent rings threaded by an Aharonov-Bohm flux are known to exhibit non-dissipative, persistent currents.   This is true of both non-superconducting (normal) mesoscopic rings \cite{ButtikerImryLandauer,Levy1990,Chandrasekhar1991,Mailly1993} and hybrid Normal metal - Superconductor (NS) rings \cite{BardeenJohnson,Waldram1975}. In both cases the thermodynamic non-dissipative   current results from the phase sensitivity of the system\rq{}s eigenenergies. The profound analogy  between the two systems has been noted since the early predictions of persistent currents in normal rings \cite{ButtikerKlapwijk}. It stems from  phase-dependent boundary conditions that induce a phase-dependent spectrum.  
The phase $\varphi$  is linked to the Aharonov-Bohm flux  $ \Phi $ via   $\varphi= -2\pi \frac{\Phi}{\Phi_0}$,  where  $\Phi_0$ is the normal flux quantum $h/e$ in the case of a  pure normal ring, and is the superconducting flux quantum $h/2e$ in the case of an NS hybrid ring. The current-phase relation at equilibrium has been measured by applying a static flux, corresponding to a dc phase bias \cite{Levy1990,Chandrasekhar1991,Waldram1975,DellaRocca2007,Fuechsle2009}. In contrast,  the  investigation of the dynamics of these systems is a more recent experimental and theoretical endeavor \cite{Deblock2002,Chiodi2011,Dassonneville2013,Virtanen2011,Ferrier2013,Sticlet2014}. The tool of choice is the measurement of  the  magnetic susceptibility $\chi= \delta I_{ac}/\delta \Phi_{ac}$ which relates the current response $\delta I_{ac}$  to a \textit{time-dependent} ac flux excitation $ \delta \Phi_{ac} \exp -i\omega t$.

The onset of dissipation in a phase-driven system, measured by the real part G of its admittance $Y=\chi/i\omega$, is still an open question. There have been two principal theoretical approaches to the problem of the conductance of a phase coherent sample. The first is the Kubo approach  derived from the calculation of the linear response of a conductor to an  ac electromagnetic field. In this approach, irreversibility is brought about by  coupling the electronic system to  a thermal reservoir with a large number of degrees of freedom (e.g. a phonon bath). The second is the Landauer-B\"uttiker approach in which the conductance of a mesoscopic system connected to electron  reservoirs   is equal to the transmission coefficient. The equivalence  between these two definitions of the conductance has been demonstrated \cite{Fisher1981} in the case of a  voltage-biased system with a continuous spectrum. In that case the conductance is proportional to the elastic scattering time with small quantum corrections related to the phase coherence time. However it has been shown that this equivalence is not valid anymore in a discrete spectrum system \cite{Reulet1994}. What happens then when an electronic system is isolated from incoherent reservoirs, either with the use of a ring  without contacts  or via  superconducting contacts ? 
\\
   Landauer and B\"uttiker addressed this basic question in a pioneering work \cite{Buttiker1986,ButtikerAnnPhys} considering a loop geometry.   They predicted that  the   response to an ac flux should contain a dissipative component  of admittance due to the delayed relaxation of populations back to the instantaneous equilibrium value. This  flux-dependent ac conductance $G(\Phi)$ is proportional to the inelastic scattering time, contrary to the Drude conductance of a connected system  that is related to the elastic scattering time. 
\\
This prediction, made in the context of mesoscopic normal metallic rings, could not be observed experimentally even in $10^5$ rings  \cite{Reulet1994_Thesis, Deblock2001_Thesis} due to the smallness of the  persistent current in  normal diffusive mesoscopic rings. 
 In the present paper we focus on the case of  a single hybrid  NS rings,  where this fundamental difference between voltage biasing  with reservoirs   and flux biasing in a ring geometry  persists in the limit of systems with a continuous spectrum. The signature of this phase-dependent dissipation is much greater thanks to  the spectral correlations borne from the superconducting boundary conditions. This allows us to reveal the full phase and frequency dependence explored in this article. 

In addition to addressing the basic question of the conductance of an isolated electronic system, our work sheds new light on the physics of a large number of Andreev Bound States, microscopic degrees of freedom of Josephson junctions that can be used to perform quantum computation \cite{Desposito2001,Zazunov2003,Janvier2015}.

\begin{figure}[h!] 
	\centering
		\includegraphics[width=0.8\linewidth]{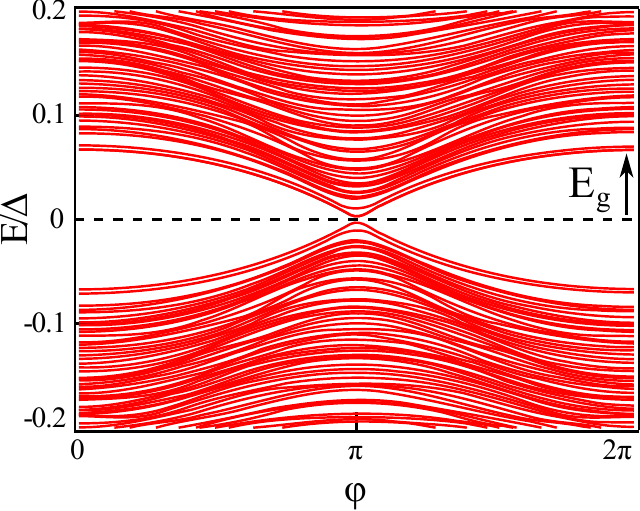}
	\caption{\textbf{Typical Andreev spectrum of a long SNS junction in the diffusive regime}. Note the phase dependence and the 2$ \pi $-periodicity of the spectrum, and the opening of the minigap. Spectrum is from \cite{Ferrier2013}.}
	\label{fig:minigap}
\end{figure}

Due to the energy gap $\Delta$ in the superconducting electrodes, a  low energy electron (hole) is retro-reflected into a hole (electron) at the interface between the N and S metals in a process called Andreev reflection. This process leads to the emergence of Andreev bound states (ABS), superpositions of electron and hole states confined in the N part of the junction  at energies below the superconducting gap. These states were observed by spectroscopy experiments  \cite{Pillet2010,Bretheau2013,Bretheau2013a,Woerkom} in junctions with few channels. They are phase-dependent and carry the Josephson supercurrent  $I_J (\varphi)=\sum f_n(\varphi) i_n(\varphi) $  where  $ f_n $ is the Fermi distribution function at energy $ \epsilon_n (\varphi) $ and the state of energy $\epsilon_n$ carries a current $i_n=-\frac{2\pi }{\Phi_0} \partial{\epsilon_n}/\partial{\varphi}$.
More specifically, we focus on the so-called long diffusive junction limit where the length $l$ of the junction is greater than the superconducting coherence length in the normal metal $\xi=\sqrt{\hbar D/\Delta}$,  with $D$ the diffusion coefficient in the normal metal.  The Andreev spectrum exhibits a phase-dependent  minigap as shown in fig.\ref{fig:minigap}.  
The  minigap only depends on the properties of the  normal metal since its amplitude at  $\varphi=0$ is $E_g=3.1\Et$ where the Thouless energy $\Et=\hbar/\tau_D$ is the energy associated with  the diffusion time $\tau_D=l^2/D$  \cite{Heikkila2002,Zhou1998,Ivanov2002}. As  measured by tunnel spectroscopy \cite{LeSueur2008}, in the continuous spectrum  limit, the minigap closes at $\varphi=\pi$  and reads:
	\begin{equation}
		E_g(\varphi)=E_g |\cos(\varphi/2)|.
		\label{eq:minigap}
	\end{equation}
 
We are interested in probing the system close to its thermodynamic equilibrium. We have therefore designed  a linear response experiment in which   one measures both the non-dissipative \mXp{} and dissipative \mXs{} current response functions  of an NS ring  \cite{Chiodi2011,Dassonneville2013}.  We measure this  response  by inductively coupling an NS ring to a multimode superconducting resonator, thereby implementing both an ac phase bias and an ac current detection at the resonator's eigen-frequencies.

  We find that at low frequency the non-dissipative current response \mXp corresponds, as expected, to the phase derivative of the supercurrent flowing through the ring, thus revealing the current-phase relation. A more striking finding is the existence of a dissipative response \mXs , revealing that the supercurrent exhibits thermal noise, as discussed in \cite{Dassonneville2013}. In that paper we analyzed the low frequency, high temperature dissipation induced by the thermal relaxation of the population of the Andreev levels and a good agreement was found with theoretical predictions obtained by solving the time dependent Keldysh-Usadel equations \cite{Zhou1997, Virtanen2011}. 
  Time dependent Usadel equations  \cite{Zhou1997, Virtanen2011,Tikhonov2015} as well as numerical simulations \cite{Ferrier2013} also predict another dissipation mechanism related to microwave-induced transitions within the Andreev spectrum. This second contribution being dominant at high frequency and low temperature showing up as an absorption peak at $\pi$  increasing as one goes deeper in the quantum regime ($\hbar \omega > E_g > k_BT$). Even if some indications of this second type of dissipation were found in \cite{Virtanen2011}  in the low temperature regime, the experiments were difficult to analyze because of screening effects.
 In this paper we present a  complete quantitative analysis of experimental  data over a wide  frequency and temperature range  on two different samples and  we show how dissipation moves from one type to the other,  in good agreement with  these  numerical and theoretical results \cite{Ferrier2013,Tikhonov2015}. 

The paper is organized as follows: in  the first section  we describe the experimental setup used to probe the dynamics of Andreev bound states over a wide frequency range. In the second section we give an overview of our experimental results.
 The complete analysis is presented in the third section.

\section{I. Experimental setup}
The experiment consists in inductively coupling an NS ring to a multimode  $\lambda/4$ strip-line superconducting resonator operating between 190 MHz and 16 GHz. The aim  is to determine the  complex magnetic susceptibility $\chi(\varphi,\omega,T)=\chi'(\varphi,\omega,T)+i\chi''(\varphi,\omega,T)$  of the ring  which relates the ac current response  $I(\varphi,\omega,T)$ to the  ac flux through the loop $\delta \phi_{ac} \exp {-i\omega t}$ in the linear response regime. The frequency $\omega$ is restricted to  the successive resonances of the resonator $\omega_n$.  The dc superconducting phase difference $ \varphi $ at the boundaries of the N wire is imposed by a magnetic flux $ \Phi $ created by a magnetic field  perpendicular to the ring plane. The ac flux is generated by the ac current in the resonator.   
 A similar technique based on single mode resonators was already used for the investigation of short Josephson junctions embedded in superconducting rings \cite{Rifkin1976,Busch2005,Troeman2008}. More related to the present work, the impedance of an array of SNS dc SQUIDS was measured using a superconducting multimode resonator \cite{finlandais}. This work focused on the temperature dependence of the effective inductive and resistive components of the SNS SQUIDs whereas our work focuses instead of the phase dependence which bares most of the signature of the dynamics of Andreev states \cite{Ferrier2013, Tikhonov2015}. 

\subsubsection*{Principle of the experiment}

The magnetic susceptibility of the NS ring $ \chi $  modifies the inductance $ L_r $ of the resonator and thus the eigen-frequencies $\omega_{n}=(2n+1)\sqrt{1/L_rC}$ and quality factors $Q_n$. 
The  induced variations of resonance frequency and quality factor  are related to the  variations of the real and imaginary components   $ \Delta L_r' $ and  $ \Delta L_r'' $  of $ L_r $  according to   $\frac{\Delta \omega_n}{\omega_n} =-\frac{\Delta L_r'}{2L_r}$ and   $\Delta(\frac{1}{Q_n})=\frac{\Delta L_r''}{L_r}$. 

\begin{figure}[h!] 
	\centering
	\includegraphics[width=\linewidth]{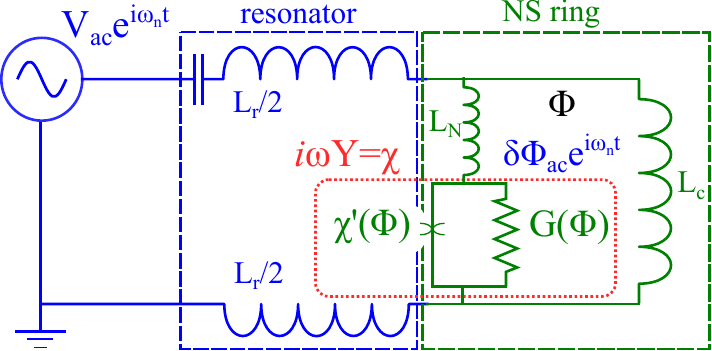}
	\caption{\textbf{Schematic representation of the NS ring inserted in the resonator}. The resonator has an inductance $L_r$.  $L_c$ and $L_N$ are respectively the coupling inductance and geometrical inductance of the normal wire,  the total geometrical inductance of the NS ring being $L_l=L_c + L_N$.  $G(\Phi)=\chi''(\Phi)/\omega$ is the dissipative component of  the admittance of the NS ring, $\chi'(\Phi)$ is its non-dissipative part. }
	\label{fig2}
\end{figure}

This leads to the following  relations between the  measured susceptibility $ \chi_m=\chi'_m+i\chi''_m $  and the perturbation of the  resonator's eigenmodes $\Delta \omega_n$ and $\Delta(\frac{1}{Q_n})$:
\begin{eqnarray}
\chi'_m=-2 \frac{L_r}{L_c^2}\frac{\Delta \omega_n}{\omega_n}
\label{chim}\\
 \chi''_m=\frac{L_r}{L_c^2} \Delta(\frac{1}{Q_n}) 
 \label{chis}
 \end{eqnarray}
$ L_c $ is the coupling inductance which is the part of the loop in parallel with the SNS junction, see fig.\ref{fig2}.
  Due to screening of the applied flux by the finite geometrical inductance of the loop $L_l = L_c + L_N $,  complex susceptibilities $\chi$ and $\chi_m$ are related through $\chi_m= \chi/(1-L_l\chi)$. The measured susceptibilities $\chi'_m$ and $ \chi''_m $  are therefore related to the intrinsic susceptibilities $ \chi' $ and $ \chi'' $ according to:

  \begin{eqnarray}
\chi'_m= \frac{\Xp}{(1 - L_l\Xp)} + \frac{ L_l\Xs^2} {(1-L_l\Xp)^2+(L_l\Xs)^2}
\label{eq:df_Xp}
\\
\chi''_m= \frac{\Xs}{(1-L_l\Xp)^2+(L_l\Xs)^2}
\label{eq:dQ_Xs}
\end{eqnarray}

When both $ L_l\chi' \ll 1 $ and $ L_l\chi'' \ll 1 $, these flux screening corrections are negligible, so that $ \chi'_m=\Xp $ and $ \chi''_m=\Xs $.
 Outside of this regime, \cite{Chiodi2011,Bastien_Thesis}, screening leads to a  dc flux rescaling and  to hysteresis at low temperature when the parameter $\beta (T) = L_l \chi_J(T)\geq 1$. We have investigated  a range of parameters where both $L_l\chi'$  and $ L_l\chi''$ are  smaller than 1 but not necessary very small. In the latter case, these screening corrections  must be  properly taken into account in order to compare our results to theoretical predictions. We also note that   the analysis of the dissipative response is more delicate than   the non-dissipative one since  $\chi ''$ is  mixed  with $ \chi'$ within first order of $L_l\chi' $. 

These screening effects have led us to design  samples  of different size which enable the exploration of both   regimes of low temperature and high frequency as well as the opposite regime of high temperature and low frequency. Experimentally we have no access to the phase independent components  of    $\chi_m$  which would require a very accurate comparison of the resonances with and without the sample. We measure instead accurately  the phase dependence of $\chi'_m$ and $\chi''_m $.  Screening  corrections  on  the phase dependent component of $\chi''$  increase drastically with frequency even at high temperature for which  $L_l \chi' \ll 1$. As discussed in the following, this  leads to a  dependence in the measured dissipative response  entirely due to screening denoted $ \chi''_s $.     We will show that we can  take advantage of this  spurious phase dependence to determine   the phase independent  value of $\chi''$  which cannot be determined otherwise.

\subsubsection*{Sample fabrication}
\begin{figure}[h!]
	\centering
	\includegraphics[width=\linewidth]{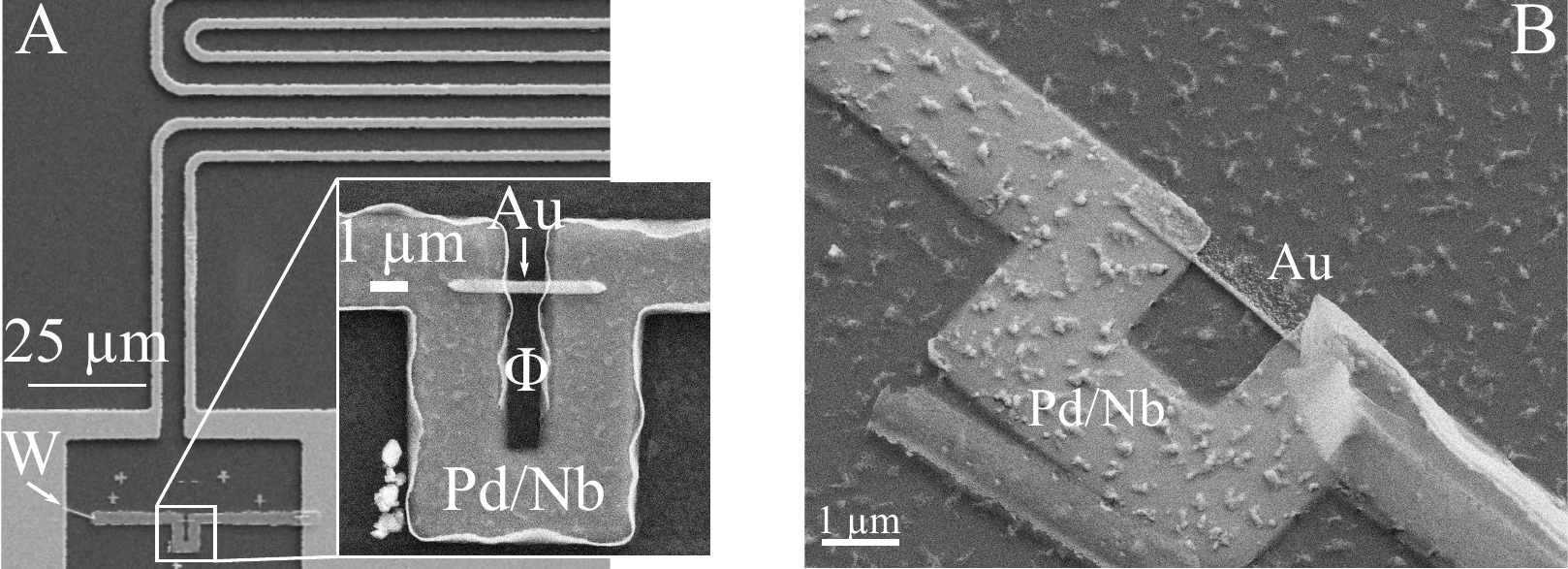}
	\caption{\textbf{The linear response is measured by inserting an NS ring in a resonator.} \textbf{Left:}  An NS ring is inserted in the middle of a $ \lambda/4 $ multimode resonator. A FIB-deposited W wire ensure a good connection between the ring and the resonator. The insert shows a close-up view of sample A. \textbf{Right:} Micrograph of sample B. }
	\label{fig:samples}
\end{figure}

The linear response is measured by inserting an NS ring in a resonator (see fig.\ref{fig:samples}).
 The resonator consists of two parallel superconducting Nb meander lines ($1\ \mathrm{\mu m}$ thick, $2 \ \mathrm{\mu m}$ wide, $ 20\ \mathrm{cm}$ long, and $4\ \mathrm{\mu m}$ apart) patterned on a sapphire substrate.   The NS ring  connects the two lines at one end of the resonator, turning it into a $\lambda/4$ line with a fundamental frequency of 190 MHz, and harmonics 380 MHz apart. One of these lines is weakly coupled to a RF generator via a small on-chip capacitance whose value is adjusted in order to preserve the high Q of the resonances (of the order of 10000), the other one is grounded. The high Q factors enable detection of  variations as small as $10^{-8}$, thus providing very accurate ac impedance measurements of mesoscopic objects.
To make the NS rings, an Au wire is first fabricated by e-beam lithography and deposition of high purity gold 99.9999 \% (Sample A), 99.999 \% (Sample B). Since sapphire is an insulating substrate, we get rid of charging effects using a conducting espacer (300Z from Showa Denko Europe GmbH) over the usual PMMA/MAA resist. The S part is  deposited in a second  alignment step  by sputtering of a Pd/Nb bilayer (6 nm Pd, 100 nm Nb). The ring is connected to the Nb resonator in a subsequent step, using ion-beam assisted deposition of a tungsten wire in a focused ion beam (FIB) microscope. This process creates a good superconducting contact between the resonator and  the Pd/Nb part of the ring. The 6 nm-thick Pd buffer layer ensures a good transparency at the NS interface, as demonstrated by the amplitude of the normal state conductance and  critical current measured with dc transport measurements on  control SNS junctions which have the same geometric properties and are fabricated simultaneously. 
Sample A has a critical current and  loop inductance which are much larger than those of sample B. It  is adequate for the investigation of the high temperature and low frequency regimes but not the low temperature and high frequency regime because of screening effects. In contrast sample B has a much smaller critical current and loop inductance and thus has negligible screening effects. This sample is  therefore more adapted for the high frequency and low temperature regimes.  For  sample A, the Au wire is  0.3~micron wide, 50~nm thick and  with a 1 micron long part that is not covered with Pd/Nb. The normal state resistance measured on a co-evaporated control sample is  $1 \Omega$. The loop inductance is estimated to be $L_l=10\pm1$ pH.  The Au wire in sample B is  70~nm wide, 30~nm thick and 1.5 micron long. The estimated normal state resistance is 10  $\Omega$,  the loop inductance is $3\pm0.3pH$.

\section{II. Overview of the experimental response}
\begin{figure}[h!]
\centering
\includegraphics[width=\linewidth]{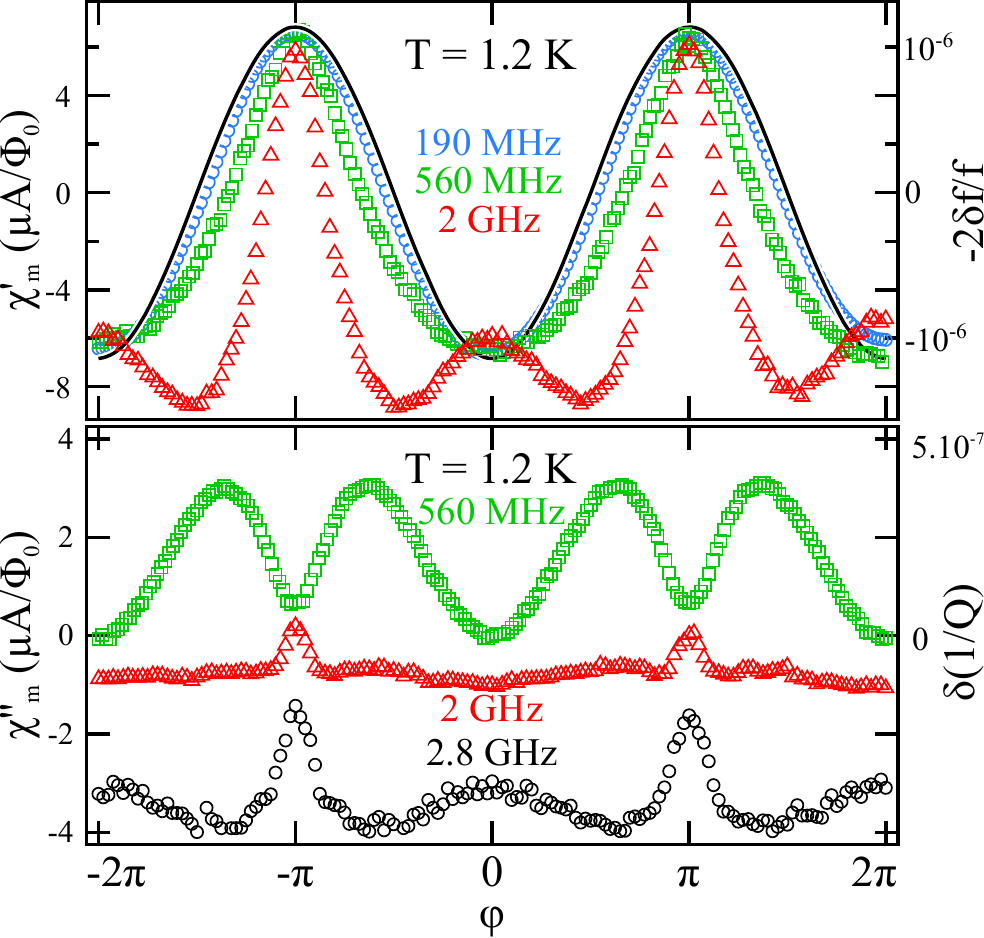}
\caption{\textbf{Evolution of the experimental response with frequency at high temperature.} \textbf{Top:} Non-dissipative $\chi'_m$ and \textbf{Bottom:} dissipative $\chi''_m$  response at several frequencies and $ T=1.2K $. Note that the amplitude between 0 and $ \pi $ of $\chi'_m$  does not depend on frequency whereas its harmonic contents does. 
For the sake of clarity, $\chi''_m(2 \rm{GHz}  ) $ and $\chi''_m (2.8\rm{GHz}$ have been arbitrarily offset.
} 
\label{fig:Xs_freq_1K2}
\end{figure}

\begin{figure*}
\centering
\includegraphics[width=1.0\textwidth]{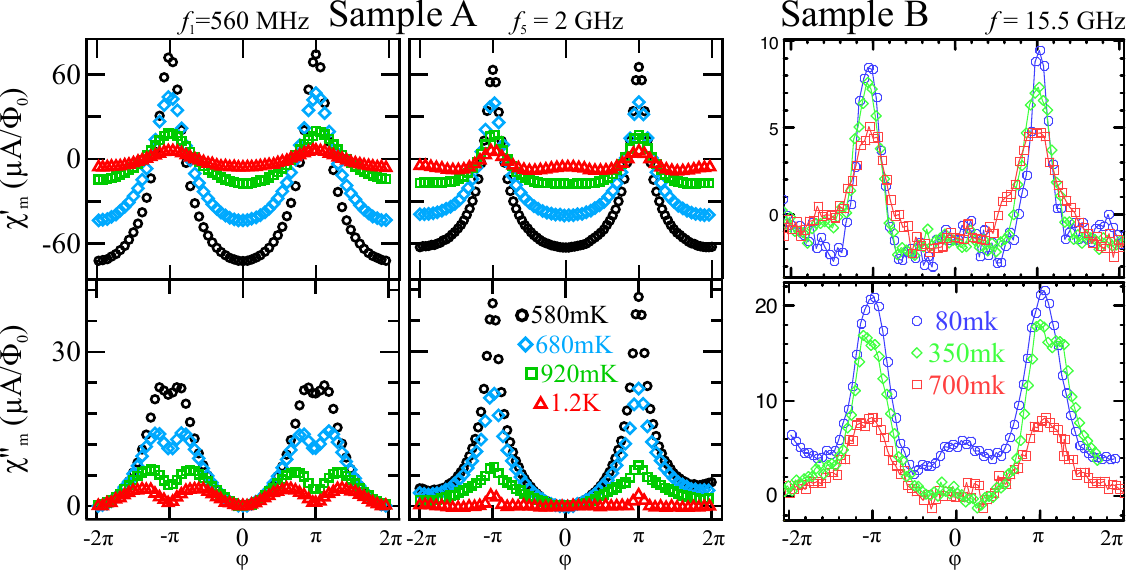}
\caption{\textbf{Evolution of the measured  response with temperature}. 
Evolution of $ \chi '_m (\varphi)$ (top) and $ \chi''_m(\varphi) $ (bottom)  with temperature. Sample A: $ f_1=560\,\rm{MHz}  <E_{Th}/h $ (left) and $ f_5=2\,\rm{GHz}  \gtrsim E_{Th}/h$ (right). Sample B: f=15.5GHz.
 Curves have been shifted vertically so that $ \chi '_m(\varphi=\pi)+\chi_m '(\varphi=0)=0 $ and $ \chi''_m(\varphi=0)=0 $.  In both cases the calibration of the susceptibility is computed from the perturbation of the resonances using expressions 2 and 3.
}
\label{fig:evo_X_T_f}
\end{figure*}

As shown in fig.\ref{fig:Xs_freq_1K2} and  \ref{fig:evo_X_T_f} we find a rich evolution of the phase dependence $ \chi_m(\varphi)$ with temperature and frequency. At the lowest accessible frequency ($ f_0=190~ \rm{MHz} $) and high temperature ($ T=1.2 \rm{K}\simeq 17\Et $),  the non-dissipative component  of $\chi_m$ measured on sample A, displays a cosine phase dependence which is the derivative of the usual Josephson current-phase dependence, sinusoidal for $ T \gtrsim E_g(\varphi=0) $ \cite{Fuechsle2009} with a decreasing amplitude   as the temperature increases.    The signal becomes highly non-sinusoidal at high frequency with a local maximum around $ \varphi=0 $  and a sharper phase dependence around $\pi$. This is a sign of  a great harmonic content with in particular an important contribution of the second harmonic. 
 Correlatively, the dissipative component $\chi''_m (\varphi)$ strongly depends on frequency.   The second harmonic clearly dominates at low frequency  and displays  a sharp dip  at $\pi$.  At high frequency instead, this phase dependence  evolves  with the emergence   a  sharp peak at  $ \pi $. The  amplitude  of this dissipation peak increases  at low temperature as shown on fig.\ref{fig:evo_X_T_f}. 
These data are the signature of large non-adiabatic contributions to the phase-dependent response of NS rings which will be analyzed in detail in the following.  On sample A,  one needs to  consider screening corrections  distorting the phase dependence of $\chi'_m$   and $\chi''_m$ at low temperature  as will be discussed in the next section.  In contrast these corrections are negligible on sample B for which one can safely assume that  $\chi = \chi_m$ over the whole temperature and frequency range investigated. Data shown in  fig.\ref{fig:evo_X_T_f} at 15.5 GHz show similar phase dependence for $\chi'$ and $\chi''$ with peaks at $\pi$ increasing at low temperature.

\section{III. Analysis}
\label{sec:analysis}

In the following we compare these results with theoretical expectations based on the linear response calculation relating the susceptibility to the phase-dependent Andreev  eigenstates  derived from the diagonalisation  of the  Bogoliubov-de Gennes Hamiltonian of a diffusive SNS junction \cite{Dassonneville2013,Ferrier2013} according to:

\begin{equation}
\begin{array}{l}
\chi(\omega,\varphi,T)=\displaystyle \frac{\partial I_J}{\partial \varphi} - \sum_{n}i_n^2
\frac{\partial f_n}{\partial\epsilon_n} \frac{i\omega }{\tau_{in}^{-1}-i\omega}-\\
\sum_{n,m \neq n} |J_{nm}|^2\displaystyle \frac{f_n  -f_m }{\epsilon_n-
\epsilon_m} \frac{i\hbar\omega}{i(\epsilon_n-\epsilon_m)-i\hbar\omega  +
\hbar\gamma_{ND}}
\end{array}
\label{eqchi}
\end{equation} 
$J_{nm}$ is the matrix element of the current operator between the Andreev 
eigenstates $ \ket{n} $ and  $ \ket{m} $ of energies $\epsilon_n(\varphi)$ and $\epsilon_m(\varphi)$, $f_n$  is the 
Fermi  Dirac function at energy $\epsilon_n$. 

\begin{figure}
	\centering
\includegraphics[width=\linewidth]{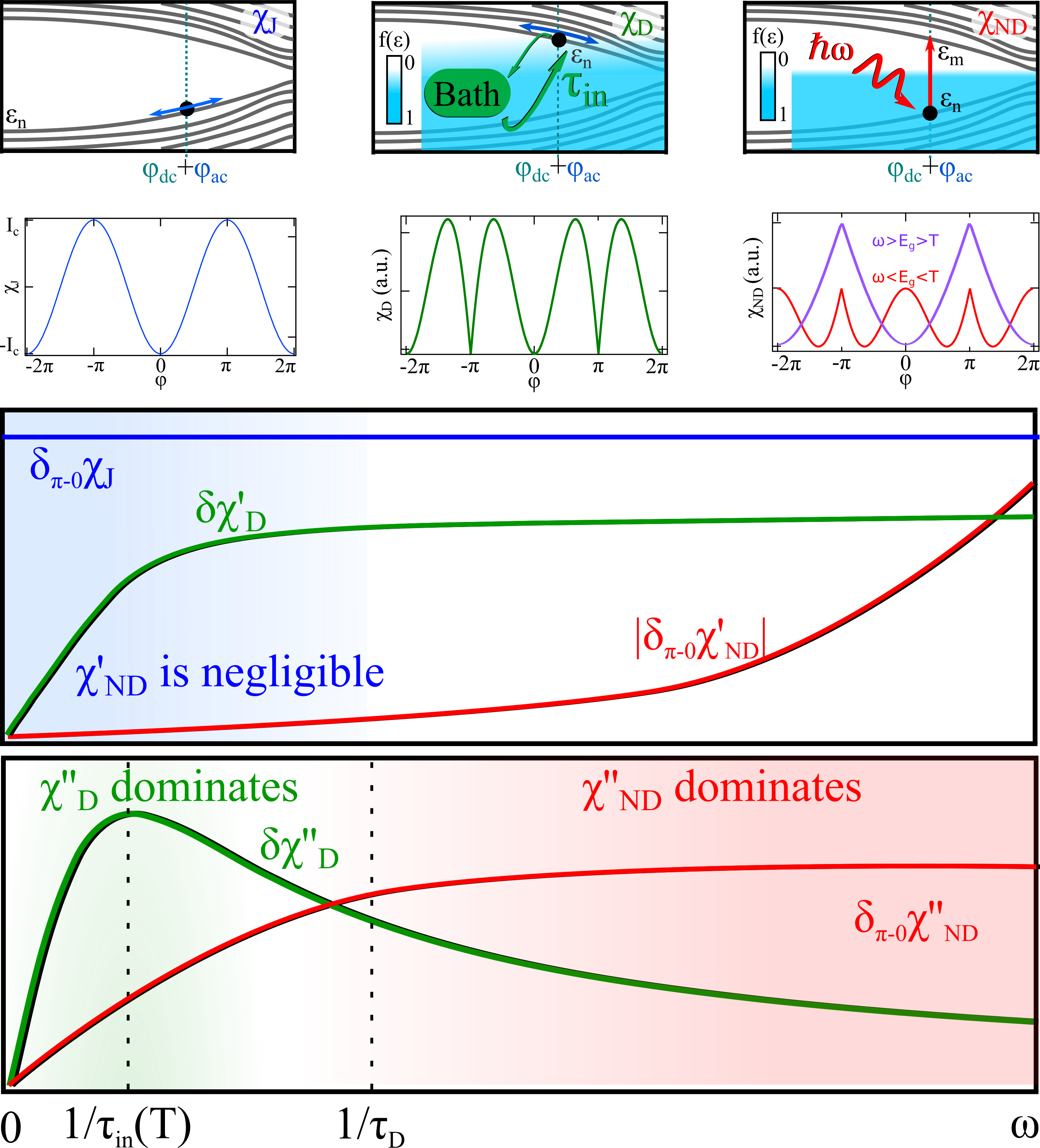}
	\caption{\textbf{Schematic frequency-evolution of the response.} \textbf{Top:} Sketch of the physical mechanisms at the origin of the finite-frequency response: (Left:)  adiabatic response $\chi_J$ ; (Middle:) relaxation of populations driven out-of-equilibrium by the finite-frequency phase-biasing, leading to $ \chi''_D $ ; (Right:) microwave induced transitions across the minigap, leading to $\chi''_{ND}$.
		\textbf{Middle:} 	Phase dependence of each contribution. $ \chi_J $ is a cosine when the equilibrium current-phase relation is purely sinusoidal, $ \chi_D $ has almost half the periodicity of  $\chi_J$. The phase dependence of $\chi''_{ND}$  depends on the temperature and frequency. At low temperature and high frequency, $\chi''_{ND}$ follows the minigap whereas at low frequency and high temperature $\chi''_{ND}$ has a phase dependence opposite to the one of $ \chi''_D $ .
		\textbf{Bottom:} Schematic frequency dependence of each contribution for the non-dissipative (Top) and dissipative (Bottom) responses. At low frequency, $\chi''_{ND}$ is negligible and $\chi''$ is dominated by $ \chi''_D $  for $ \omega\tin\sim 1 $. At high frequency $\chi''$ is dominated by $\chi''_{ND}$.}
	\label{fig:Overview_Theo}
\end{figure}

The first term is the zero frequency susceptibility 
of the ring, $\chi(\omega=0)=
\partial I_J/ \partial \varphi$. We call it Josephson contribution $ \Xj $. The second and third terms only exist at finite frequency and  describe   the  non-adiabatic, dynamical responses due respectively to 
the relaxation of the populations $\chi_D $,   and  to the transitions between 
the levels induced by microwave photons emission or absorption $\chi_{ND}$, 
the quantities $1/\tau_{in}$ and  $\gamma_{ND}$  being respectively the 
diagonal and non-diagonal relaxation rates of the system  determined by its 
interaction  with its thermodynamic environment.  Dissipation is described by the imaginary 
components of the diagonal and  non-diagonal susceptibilities $\chi''_D$ and $\chi''_{ND}$. Their different phase dependence reveal their different physical origins as sketched in fig.\ref{fig:Overview_Theo}.  $\chi''_D$ is proportional to the square of the single level current and must be zero at phases multiples of $\pi$. In contrast, $\chi''_{ND}(\varphi)$ is determined by the interplay of the phase-dependent non-diagonal elements of the current operator, the occupation of the levels, and  the presence of the minigap in the density of states. Its phase dependence therefore strongly depends on the relative amplitude of $\hbar\omega$, $k_B T$ and $E_g$. In addition, $\chi''_{ND}(\varphi)$ has a phase-independent part that is expected to ultimately gives the normal conductance when no trace of coherence is left.  In the regime of high temperature and low frequency the non-diagonal elements of the current operator lead to a phase dependence of $\chi''_{ND}(\varphi)$ opposite to $\chi''_{D}(\varphi)$ whereas at low temperature dissipation is depressed except  at $\pi$ due  to the minigap, leading to a dissipation peak  at $\pi$ in the phase-dependent part of the dissipation.  These phase and frequency dependence are discussed in more details  in  Appendix A. The analysis of our experimental results   reveals these contributions and the physical mechanisms underpinning them. We first discuss data on sample A that reveals the adiabatic response as well as the dynamic response in the low frequency and high temperature regime.   We also determine the phase-independent contribution to the conductance that is revealed through flux screening effects. In contrast, data on sample B (for which screening corrections are negligible) show unambiguously  a dissipation peak  at $ \pi $ related to the presence of the minigap. 
\subsubsection*{Notations}
	  We  use the following notations: $\delta_{ \varphi_1 -\varphi_2}\xi =\xi(\varphi_1)-\xi(\varphi_2) $ and $\delta \xi $ is the maximum  absolute amplitude of the phase dependence of $\xi$ which stands for $\chi'$, $\chi''$, $\chi_m$ or $\chi''_m$.

\subsection{Adiabatic response \mXj}
 For frequencies such that $ \omega\ll\gamma_D \leq \gamma_{ND}$, the diagonal and non-diagonal contributions are negligible, so $ \chi'(\varphi) =\X_J(\varphi)$ yields the phase derivative of the current-phase relation at equilibrium. 
The phase dependence  of $\chi' $  is obtained from $\chi'_m (\varphi)$     after correction of the self inductance effects   via:
\begin{equation}
 \delta_{\varphi-0}\chi'_m = \chi'_m(\varphi) -\chi'_m (0) = \frac{\chi'(\varphi)}{1-L_l \chi'(\varphi)}  - \frac{\chi'(0)}{1-L_l \chi'(0)} 
\end{equation}
 using that $\chi'(\pi)= - \chi'(0)= \chi_J(\pi)$.  

\begin{figure}
\centering
\includegraphics[width=0.7\linewidth]{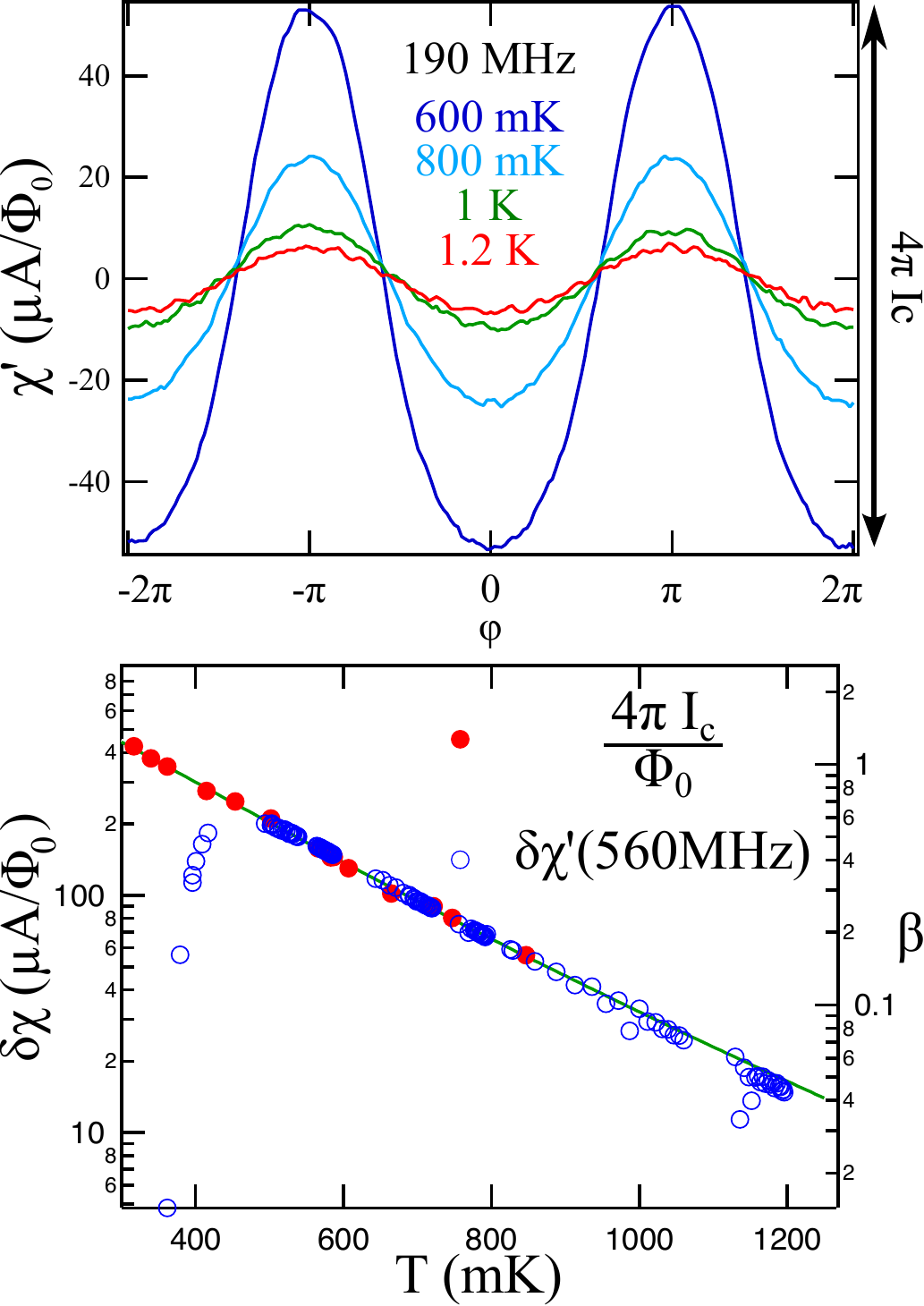} 
\caption{\textbf{Characterization of the ring at low frequency.} \textbf{Upper panel} Phase-dependence of the measured non-dissipative response $\chi'_m$ at 190~MHz  and several temperatures. At low frequency, the amplitude of $\chi'_m$ is directly proportional to the ring's critical current: $ \delta_{\pi-0} \chi' =4\pi I_{c}/\Phi_{0}$. \textbf{Lower panel} Temperature dependence in semi-log scale of the amplitude of the non-dissipative response of the ring at 560~MHz, $ \delta \chi'(f=560\rm{MHz}  ) $   (circles) and of the control sample using that $\delta \chi'= 4\pi I_{c}/\Phi_{0} $ (full circles) along with its fit (solid line). $ E_{Th}=71 mK $ is found. 
}
\label{fig:characterisation}
\end{figure}

For  the lowest eigen-frequency of the resonator,  see fig.\ref{fig:characterisation}, $\chi'(\varphi)$ is well described by a cosine and its amplitude 
is  simply related to the critical current by $\delta_{\pi-0}\chi'=\dfrac{4\pi}{\Phi_0} I_c= 2\chi_J(\pi)$. Noting that \mXpd{} does not modify $\delta_{\pi-0}\chi'$, we find (see fig.\ref{fig:characterisation}) that this relation between the amplitude of \mXp and $I_{c}$ holds in the whole temperature range investigated where  $k_BT \gg E_{Th}$.  $ \delta_{\pi-0} \chi' (T) $ follows   the expected exponential decay of the Josephson critical current   $ I_c(T) = I_J(0)\exp(-k_BT/3.6 E_{Th})$ \cite{DubosJLT}. Fitting this dependence yields $ E_{Th}/k_{B}= 71\pm 2 $ mK.  These results agree with the temperature dependence of the switching current of a current-biased control junction that has similar geometrical characteristics. The analysis of the  amplitude of $\delta_{\pi-0} \chi'$ in the low frequency regime thus allows to perfectly characterize the junction and yields the total minigap width $2E_g(0)/h= 9~\rm{GHz}$   and the corresponding diffusion time across the junction $\tau_D= 0.1$ ns.

\subsection{ Main contribution at low frequency  and high temperature: 
 relaxation of the  populations of Andreev states.}

In this paragraph  we focus on  the diagonal susceptibility which is due to the  Debye-like relaxation of  the occupation of the Andreev levels  driven out-of-equilibrium by the ac phase excitation \cite{Zhou1997}. This is the first non-adiabatic contribution observed at modererate frequencies when frequency is of the order of the relaxation time of the phase-dependent populations $ \tin $.  According  to expression \ref{eqchi}, frequency and phase dependence  simply factorize. The frequency dependence is described by the simple  Debye relaxation law in $\omega\tau_{in}/(1-i\omega\tau_{in})$ and the phase dependence of $\chi_D$  is therefore  identical for both its real and imaginary components.

\begin{figure}[h!]
\centering
\includegraphics[width=1.0\linewidth]{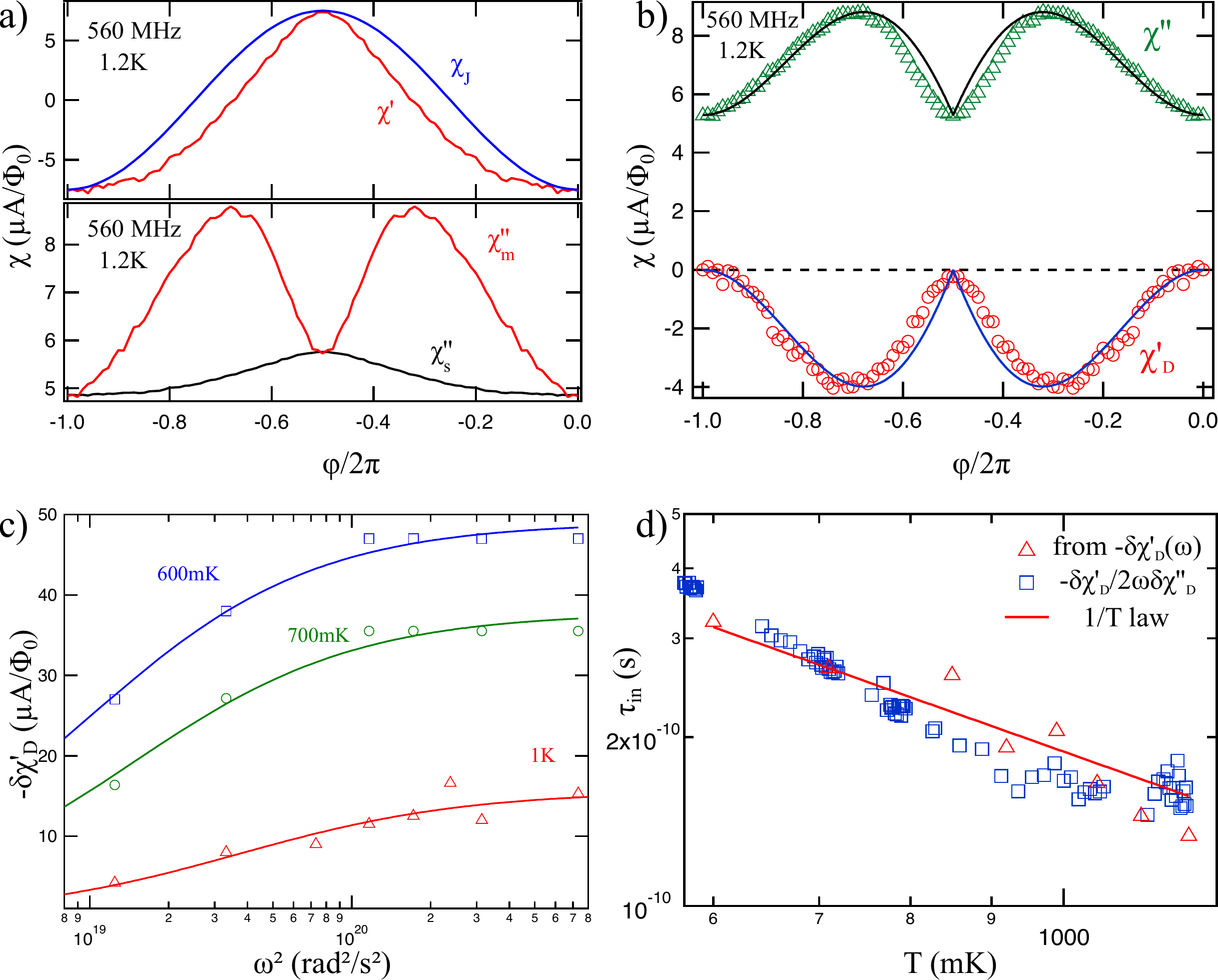}
\caption{\textbf{Diagonal contribution to \mXp and \mXs.}
\textbf{a:} Phase dependence at $ T=1.2\,\rm{K} $ and 560\,MHz of:  the adiabatic response $ \chi_{J} $ and the finite frequency response $ \chi' $ (top panel), the measured dissipative response  $ \chi''_m $ and the screening contribution $ \chi''_s $ (bottom panel). We note that  screening effects are negligible on $ \chi' $ and therefore $ \chi'_m\simeq \chi' $ at this temperature.
{\bf b: }  At high temperature and for  $ f \lesssim E_{Th}/h $, experimental $\chi'_D=\chi'-\chi_J $ and $ \chi"= \chi''_m- \chi''_s $ are found to be in very good agreement (within a factor 2) with theoretical prediction of  Usadel equations  given by the Lempitski function $ F_{U} $ (solid lines).  Theoretical predictions have been multiplied by an arbitrary factor so that their amplitude matches  the experimental ones.
{\bf c: }Frequency dependence of  $ -\delta \chi'_D$ :  the maximum  of $-\chi'_D(\varphi)$, at different temperatures (symbols) compared to the theoretical prediction (see main text). 
{\bf d: } Temperature dependence of $ \tau_{in} $ independently determined from the frequency dependence of  $ -\delta \chi'_D$ (triangles) and from the ratio  $ \delta \chi'_D/(\omega\delta \chi''_D)$ at 560\,MHz (squares) along with a $ T^{-1} $ law (solid line).
}
\label{fig:phase_dep_limp}
\end{figure}

At these low enough frequencies so that the non-diagonal contribution to \mXp{} stays negligible,
the most reliable way to  access this phase dependence at temperatures above the Thouless energy  is  to  subtract the adiabatic response to the   susceptibility:  $\chi'_D (\varphi)= \chi'(\varphi) - \chi_J(\varphi) $ (see fig.\ref{fig:phase_dep_limp}.a) where $\chi_J(\varphi)$ is a simple cosine. The experimental  determination  of  $\chi''_D (\varphi) $  is more  delicate  and needs to be restricted to the range of temperature and frequency where  both screening $ \chi''_s $ and $\chi''_{ND}$ contributions are negligible.  An example is shown in fig.\ref{fig:phase_dep_limp} 
 at high temperature (T=1.2 K) and low frequency (f=560 MHz) where the   measured signal $ \chi''_m$ is clearly dominated by  $\chi''_D (\varphi)$. 
Indeed the phase  dependence exhibits the characteristic shape of $  \sum  \partial f/\partial \epsilon_n $, typical of the relaxation of Andreev levels occupations. It is zero at 0 and $\pi$ as  expected for the square of the single level  current  with a sharp dip at $\pi$ coming from the contribution of the Andreev levels close to the minigap \cite{Dassonneville2013}. We compare  successfully this   flux dependence  with  the Lempitski function  $F_U(\varphi)$ \cite{Lempitskii1983}  corresponding to  the  time-dependent Usadel equations (see Appendix A and \cite{Virtanen2011}). We  also observed that this shape of  $ \chi'_D(\varphi)$ is independent of temperature   \cite{Dassonneville2013} for $   k_BT \gg E_g$. The  frequency dependence of $\delta\chi'_D$ are shown in fig.\ref{fig:phase_dep_limp} for different temperatures.  They can be  fitted by the expected $\frac{ (\omega\tau_{in})^2}{1+(\omega\tau_{in})^2}$  which enables to determine the characteristic time $\tau_{in} (T)$.

We find values of $\tau_{in}$  varying between  like $1/T$ \cite{compare2013}. This temperature dependence is similar to the predicted  T variation of the phase coherence time in a  short diffusive wire when it is limited by electron-electron  inelastic scattering \cite{Blanter1997,Texier2005,Ferrier2008,Capron2013}. However the prefactor in our experiment is more than  2 orders of magnitude smaller.  This result points towards a different mechanism, possibly due to the  thin Pd layer at the interface between the Nb and Au yielding low energy sub-gap excitations.

\begin{figure}[h!]
\centering
\includegraphics[width=\linewidth]{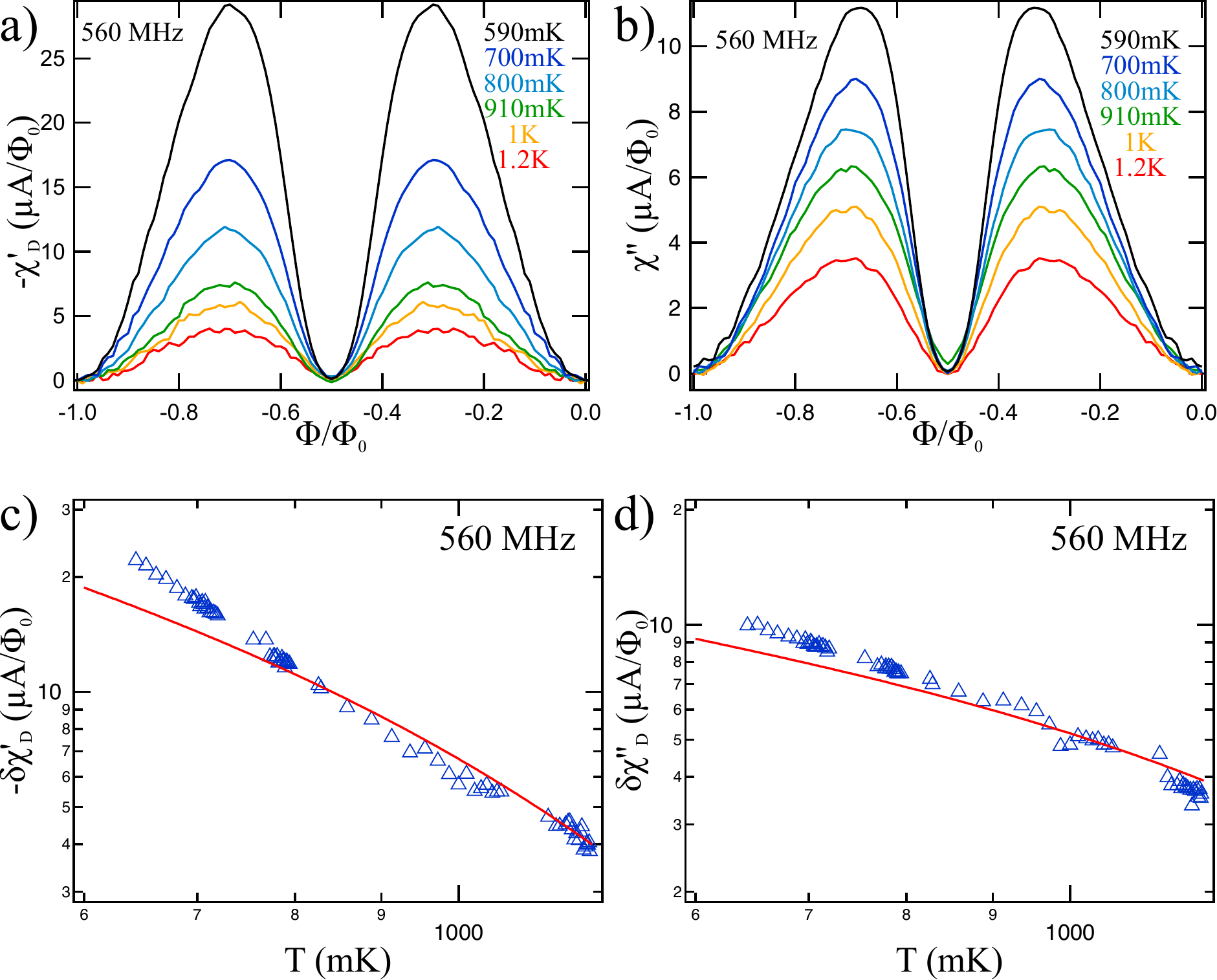}
\caption{\textbf{Evolution of the diagonal contribution with temperature.} \textbf{Top:}
-\mXpd{} and $ \chi''$ at $ f=560~\rm{MHz} $ and several temperatures. As seen from the phase dependence,  $ \chi''$ is dominated by $ \chi''_D$ at this frequency. \textbf{Bottom:} Temperature dependence of the amplitude of -\mXpd (resp. $ \delta\chi''_D $)  at several frequencies (symbols) along with  the corresponding $T^{-1}\frac{\omega^2\tau_{in}^2}{1+\omega^2\tau_{in}^2}$ temperature dependence (resp. $T^{-1}\frac{\omega\tau_{in}}{1+\omega^2\tau_{in}^2}$) (solid lines). $\frac{\omega^2\tau_{in}^2}{1+\omega^2\tau_{in}^2}$ and $\frac{\omega\tau_{in}}{1+\omega^2\tau_{in}^2}$ are determined using the $ T^{-1} $ fit of the experimental $ \tin(T) $ (fig. \ref{fig:phase_dep_limp})
}
\label{fig:Xpd_Tdep}
\end{figure}

We now return to the  temperature dependence of  $ \chi'_D $  resp.  $ \chi''_D $ shown in fig.\ref{fig:Xpd_Tdep} at fixed frequency. They  follow the   expected $ T^{-1}\frac{\omega^2\tau_{in}^2}{1+\omega^2\tau_{in}^2}$ (resp. $T^{-1}\frac{\omega\tau_{in}}{1+\omega^2\tau_{in}^2}$). The $1/T$ prefactor comes from the energy derivative of the Fermi  distribution at $T\gg E_g$. Interestingly, this temperature dependence is much slower than the exponential decrease of the supercurrent. This result agrees with the predictions of Zhou and Spivak on the conductance of a SNS junction \cite{Zhou1997}. We note that a similar temperature dependence  was also found for the fractional Shapiro steps  also due to this relaxation of the population of the Andreev states \cite{Lehnert1999,Argaman1999, Dubos2001}.


\subsection{ Main contribution at high frequency  and low temperature: 	microwave-induced transitions between Andreev states.}

 The aim  of the following paragraph is to analyse the dissipation  when frequency becomes larger than $1/2\pi \tau_{in}$ and the contribution of induced transitions within the Andreev spectrum, $\chi_{ND}$, becomes important. The analysis of $\chi'_{ND}$ is involved (see Appendix A) and we therefore focus on $\chi''_{ND}$. It can be noted however that $\chi'_{m}$ is almost temperature independent up to 700 mK at 15.5~GHz (see fig.\ref{fig:evo_X_T_f}) which 
 shows that \mXpnd{} is depressed at high frequency and low temperature and becomes much smaller than \mXj{}. The amplitude of  \mXj{} is estimated to be of the order of $ 55 \rm{\mu} A/\Phi_0 $ at $ T \lesssim \Et $ for sample B.
 
 We consider separately the high and low temperature regimes compared to the minigap which were respectively explored in sample A and sample B. 
On sample A,  screening corrections  are important and strongly distort the measured phase dependence. However these distortions can   be exploited to determine the phase-independent component of $\chi''$. This is done by the determination of $\chi''(\pi)$   which is not simply equal to $ \omega G_N$ but is  found to vary with temperature  in agreement with the predictions  of  \cite{ Ferrier2013,Tikhonov2015}. The opposite regime of low temperature was explored using sample B and reveals the signature of the minigap. The very small amplitude of the signals measured  did not allow to explore the high temperature regime on this second sample.

\subsubsection{  Temperature above the minigap: competition between $\chi''_{D}$ and $\chi''_{ND}$ }

We show that in this   regime explored on sample A, not only the phase dependence of  the non-diagonal contribution $\chi''_{ND}$   but also the absolute amplitude of $\chi'' (\pi)$ can be revealed. We start from  expression (\ref{eq:dQ_Xs}) relating the measured $\chi''_m$ to $\chi'$ and $\chi''$ in the limit where $\chi'' \ll \chi'$ :
\begin{equation}
\chi''_m= \frac{\Xs}{(1-L_l\Xp)^2}
\label{eq:dQ_Xsimple}
\end{equation}

\begin{figure}
 \centering
 \includegraphics[width=1.0\linewidth]{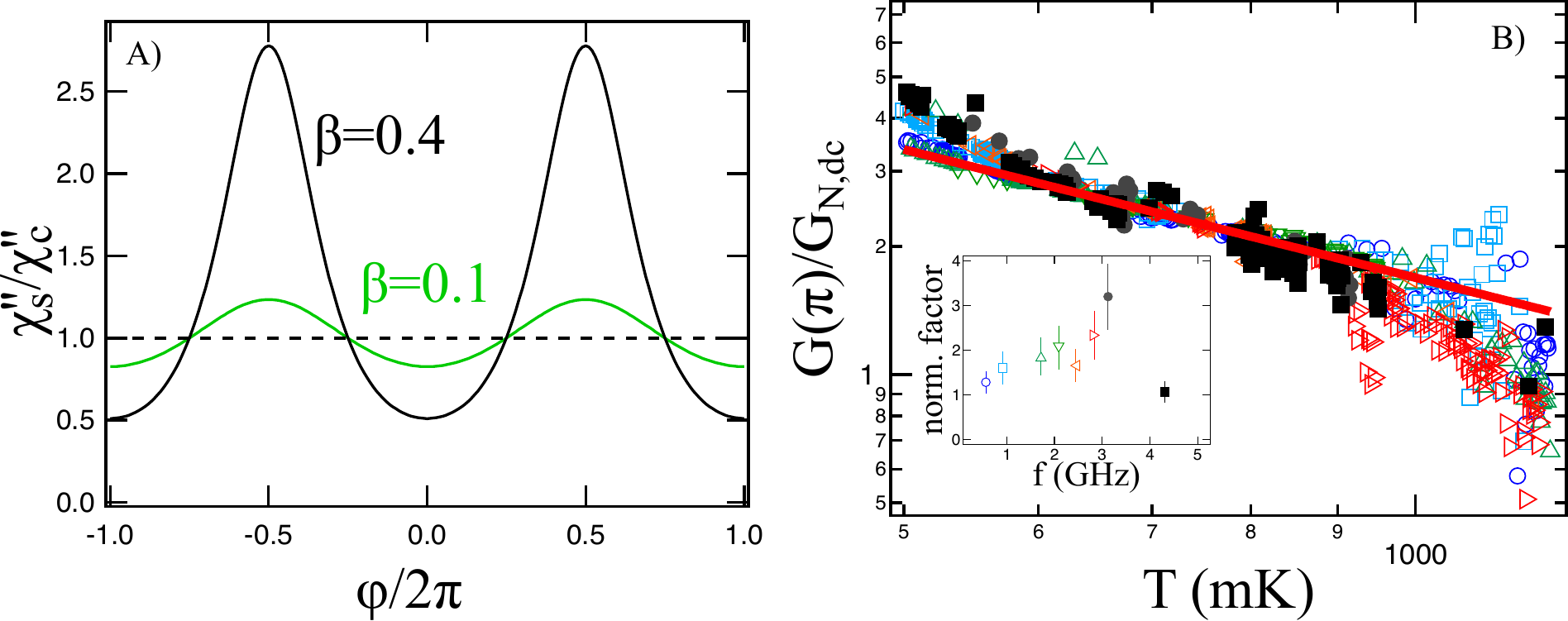}
\caption{\textbf{Determination of $ \chi''(\pi,T,\omega) $.} \textbf{A)} Phase-dependent dissipative response $ \chi''_s $ due to screening.  $ \chi''_s= \chi''_c/(1- L_l\chi'(\varphi))^2 $ is shown for $ \beta=0.4 $ (black) and $ \beta=0.1 $ (green)  where a simple  $ L_l \chi'(\varphi)=-\beta \cos(\varphi) $ has been used. \textbf{B)} The temperature dependence of $ G(\pi) $ is obtained from the ratio between the amplitudes of $ \chi''_m $ and $ \chi'_m $  as explained in Appendix B. (Different symbols correspond to the  different frequencies  shown in the inset). A good agreement with a $ 1/T $ law (solid line) is observed in a regime of intermediate temperatures. Data have been renormalized so that $ G(\pi)$ at 1.2K$\simeq G_{N,dc} $ with  $ G_{N,dc} $ the conductance of the control sample. The frequency-dependent normalization factor is shown in the inset.}
\label{fig:Xss}
\end{figure}

This equation shows that even a constant (phase independent) $\chi''= \chi''_{c} $ will display a phase dependence given by $ \chi''_s(\varphi)= \chi''_{c}/(1-L_l\Xp(\varphi))^2$ as shown in fig.\ref{fig:Xss}. This screening contribution  needs to be determined in order to deduce the intrinsic $\chi''(\varphi)$. The protocol is described in detail in Appendix B.  $\chi''(\pi)$ is extracted from the ratio $R_m=  \delta_{\pi-0} \chi''_m/ \delta_{\pi-0}\chi'_m $.
It is  interesting that we could take advantage  of screening to determine the absolute value of $\chi''$ at $\varphi=\pi$  and therefore the phase independent component of $\chi''$ which {\it cannot be determined otherwise}.  Since  the minigap closes at $ \varphi=\pi $,  with a density of states similar to the one of a normal metal, it could be  expected that     $ \chi''(\pi)$ would be given by $\omega G_N $ and  independent of temperature. As shown in fig.\ref{fig:Xss}, we find  instead  that  $ \chi''(\pi) $ {\it increases} above $G_N $ at low temperature . Within our experimental uncertainty, the value of $ G(\pi) $ above 1K  is similar to the normal state conductance of the control sample $ G_N= 0.9\pm 0.2 \rm{S} $ and  $ G(\pi,T)= \chi''(\pi,T)/\omega \propto  T^{-1} $ $ G(\pi)$ .
 This important  increase of $G(\pi)$ up to values much larger than $G_N$ at low temperatures is surprising. It is in qualitative agreement with recent predictions \cite{Virtanen2011,Ferrier2013,Tikhonov2015}. We understand this effect as  being related to phase-dependent correlations in the Andreev spectrum extending to energy scales  much larger than the minigap (see Appendix A)  with a special contribution of  matrix elements $|J_{-n,n}|^2$  coupling electron-hole symmetric Andreev levels. Moreover it can be argue that this increase in dissipation should be taken into account when considering the Joule power dissipated by a junction to determine the temperature at which hysteresis in the I-V characteristic appears as done in \cite{Courtois2008,DeCecco2016}

 This determination of $\chi''(\pi,T)$ is essential to reconstruct the intrinsic phase dependence of $\chi''$ from $\chi''_m (\varphi)$ according to:
\begin{equation}
\chi''(\varphi) =  (1-\beta(\varphi))^2\left( \chi''_m (\varphi) - \chi''_m(\pi)  + \frac{\chi''(\pi)}{(1-\chi'(\pi)L_l)^2} \right)
\end{equation}
where $\beta(\varphi)=L_l\chi'(\varphi)$.

\begin{figure}
 \includegraphics[width=1.0\linewidth]{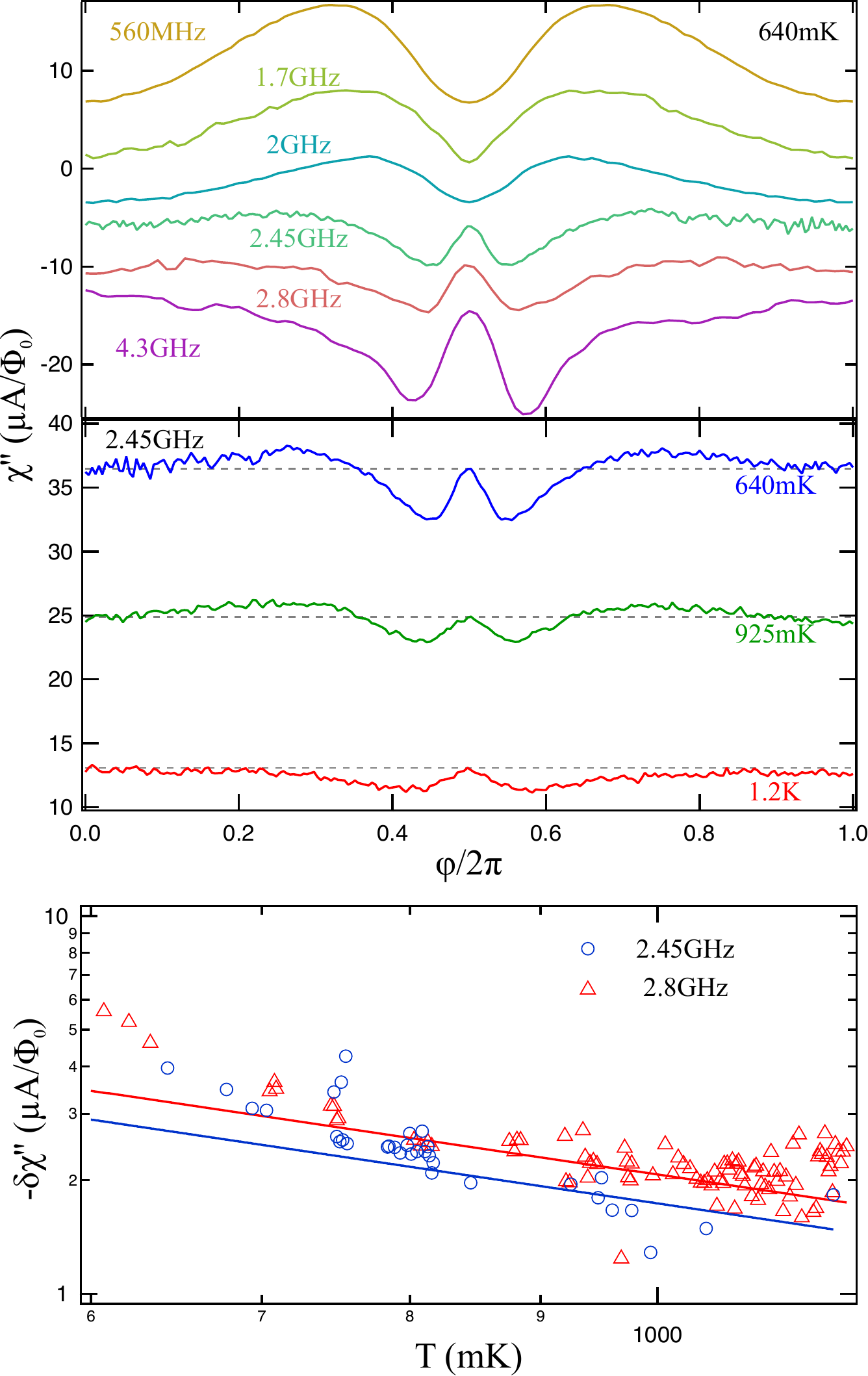}
 \caption{\textbf{Phase and temperature dependences of the dissipation at $ T>  E_g/k_B $.} \textbf{Top panel:} evolution of the phase dependence of $\chi''$ with frequency at 640mK  from top to bottom  0.56, 1.7, 2.2.45, 2.8, 4.3~GHz,(curves are arbitrarily shifted from one another). \textbf{Middle panel:} evolution of  $\chi''(\varphi)$   for different  temperatures  at 2.45~GHz, (in contrast with upper panel there is no added offset between the curves). \textbf{Bottom panel:} temperature dependence of  $\delta\chi''$,  the continuous lines corresponds to a 1/T decay.
 }
 \label{fig:XsndT}
 \end{figure}

The  resulting phase dependence of $\chi''$ are shown in fig.\ref{fig:XsndT} for different frequencies and temperatures. One can see the crossover between the low frequency regime where  $\chi''_D$    (characterized by  a minimum of dissipation at $\pi$) is dominant  and the  high frequency regime , $ \omega\tau_{in}\gg1 $, where   $\chi''_{D}$  becomes negligible  compared to  $\chi''_{ND}$, and the phase-dependent part of the dissipation starts to peak  at $\pi$ with a peak amplitude  increasing  with frequency. This evolution in the phase dependence reflects the competition between the  contribution of the diagonal elements of the current operator  (minimum at $\pi$  due to the cancellation of single level currents)  and the contribution of the non-diagonal matrix elements of the current operator   which, in contrast, are maximum at $\pi$ where Andreev levels anti-cross.  In this range of temperature,  $T\gg E_{Th}$  the contribution of the phase-dependent  minigap is  still negligible  and the  phase dependence of $\chi''_{ND}$  is essentially  determined  by the $ |J_{nm}|^2 $  and is  opposite in sign to the phase dependence of $\chi_D$ (see appendix A).


 The amplitude of  the phase-dependent modulation of $\delta \chi''_{ND}$ also follows a $ 1/T $ law like  $\chi''(\pi)$   stemming from the temperature dependence of the  Fermi functions differences $f_n-f_m$.
This 1/T dependence, similar to the temperature dependence of $\chi_D$ in the same T range ($T\gg E_g$), is much slower than the  exponential T decay of the Josephson current, demonstrating that the dynamical response is much more robust  than the critical current to temperature.

%

   \subsubsection{ Low temperature and high frequency: signature of the minigap }
   
\begin{figure}

	\includegraphics[width=\linewidth]{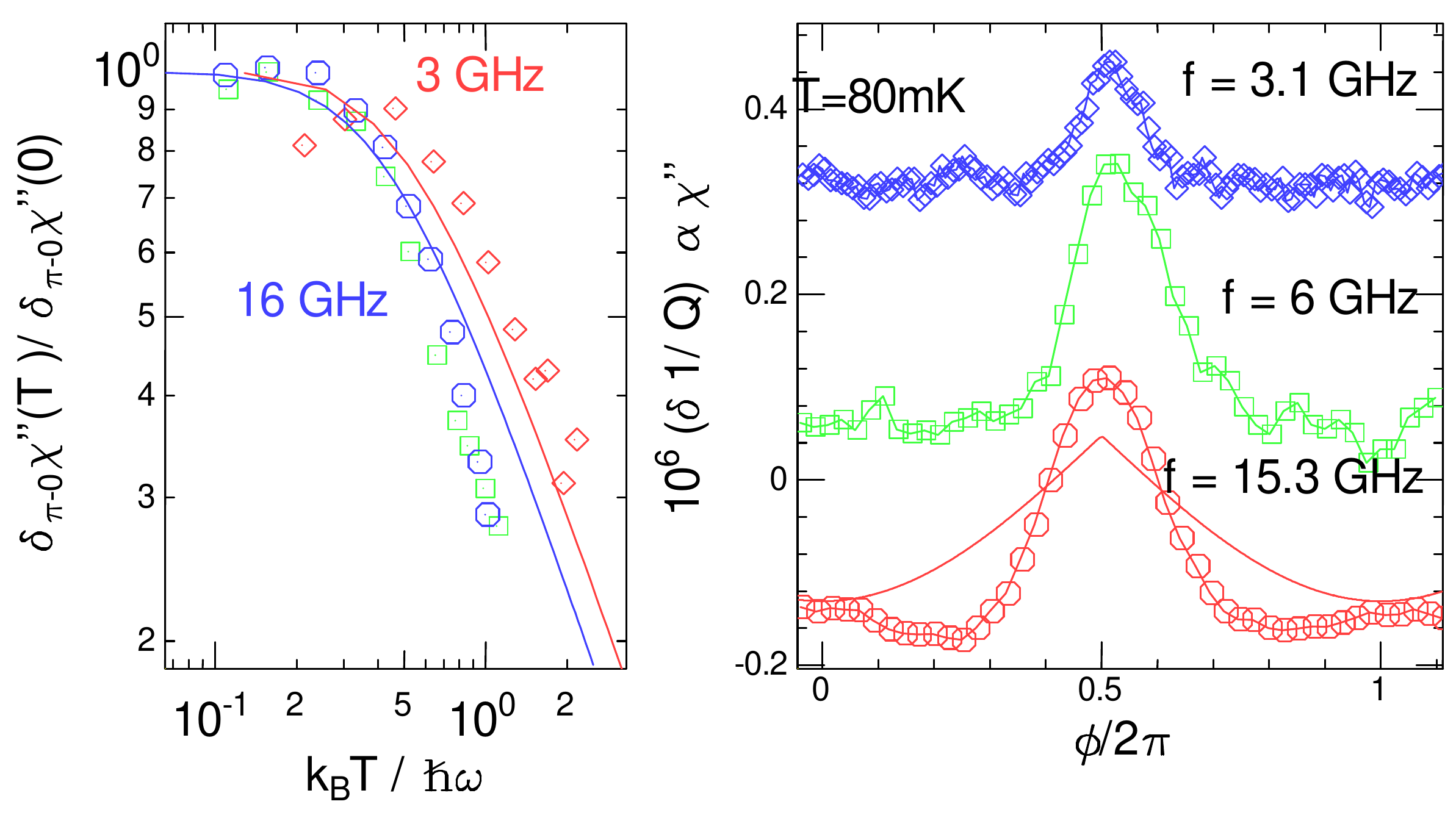}
	\caption{\textbf{Phase and temperature dependences of the dissipation at $ T<  E_g/k_B $.} \textbf{Left:} Temperature dependence of $\delta _{\pi-0} \chi''$ at 16 (squares), 15.3  (circles) and 6 GHz (diamonds, we note a saturation at temperatures corresponding to $ 0.25\hbar\omega$. The plain lines correspond to equation \ref{fitT}  valid for $\hbar\omega \gg E_g$. \textbf{Right:} Phase dependence $\chi_{ND}(\varphi)$ measured for sample B at 80mK  for different frequencies below and above the minigap corresponding to 5GHz in comparison with the phase-dependent minigap (plain line). The dissipation peak reflects closing of the minigap but the overall phase dependence has a shape which differs from the minigap, indicating a sizable contribution of the phase dependence of non-diagonal matrix elements. }
	
	\label{chisdeTetf}
\end{figure}

We now consider high frequency data measured on sample B.  Since for this sample screening corrections are completely negligible with $ \beta\leq 0.03 $, the  measured  response coincides with the intrinsic one down to very low temperature compared to the minigap. As shown in the panel B of fig.\ref{fig:evo_X_T_f} and fig.\ref{chisdeTetf} the phase dependence of $\chi''$  displays  a sharp dissipation peak   at $ \pi $.  The phase dependence of $\chi''(\varphi)$  is shown  in fig. \ref{chisdeTetf} for different frequencies of the order or larger than the minigap of sample B estimated as: $2 E_g/h=5$ GHz . This phase dependence as well as the   amplitude of the dissipation peaks at $\pi$ are found to be only slightly frequency dependent. This result is in agreement with the expected saturation  of $\delta _{\pi-0} \chi'' $  at frequencies greater than the Thouless energy close to the estimated  value of $2G_N E_g /\hbar $,  see Appendix A and \cite{Ferrier2013}. We find a saturation at  $\delta _{\pi-0} \chi'' \simeq 15\mu A/\Phi_0 $  whereas   $2G_N E_g /\hbar = 12 \mu A/\Phi_0$.

  The temperature dependence of $\delta _{\pi-0} \chi''$ is  shown fig.\ref{chisdeTetf} for different frequencies above and below $E_g$. It varies as $1/T$   and  saturates  at a temperature below $0.4K$ for 15 and 16~GHz whereas no clear saturation can be detected at 3~GHz.   Our data  are compared with the theoretical expectation neglecting the phase dependence of matrix elements of the current operator:
 \begin{equation} 
\begin{array}{l}
  \delta _{\pi-0} \chi''(T) = G_N\int_{-E_g}^{+E_g}  (f(E+\hbar\omega)-f(E))dE = \\ \frac{G_N 
  	k_BT}{\hbar} \left(-\log \frac{e^{\frac{\omega -\text{Eg}}{T}}+1}{ e^{\frac{\text{Eg}+\omega }{T}}+1} + E_g/k_BT \right)
\label{fitT}
\end{array}
  \end{equation}
	 The  temperature dependence of $\delta _{\pi-0} \chi''(T)$   is strongly related to the behavior of the Fermi function difference $ \delta(T,\omega)= f(E)-f(E+\hbar\omega)$.  $ \delta(T,\omega)$ exhibits a  E/T  dependence at high temperature compared to $\hbar \omega$ and saturates at low temperature  of the order of $\hbar \omega /4$ to a value equal to  $4E/\hbar\omega$.
 
In spite of this  reasonable agreement between our data and these predictions for the T dependence  of $\delta _{\pi-0} \chi''(T)$, $ \chi'' $ displays a phase-dependence sharper than expected, for all frequencies investigated above the minigap  (see fig. \ref{chisdeTetf}).  
	This suggests that the  contribution of the phase-dependence of 
the squared matrix elements of the current operator still 	play a role even at rather large energy ($ \hbar\omega\sim 3 E_g $). 
Including an additional contribution due to the e-h symmetric transitions with a phase-dependence peaked at $ \pi $  gives a better agreement (not shown) at the cost of introducing additional parameters. This argument is supported by numerical simulations, compare for instance \mXs(15.3~GHz, 80~mK) with fig.7 in \cite{Ferrier2013}. Such a description  requires  requires a more quantitative analysis of the energy correlations of the $ |J_{nm}|^2 $ and is is left for future studies.

\section{Conclusion}
\begin{figure}
\includegraphics[width=\linewidth]{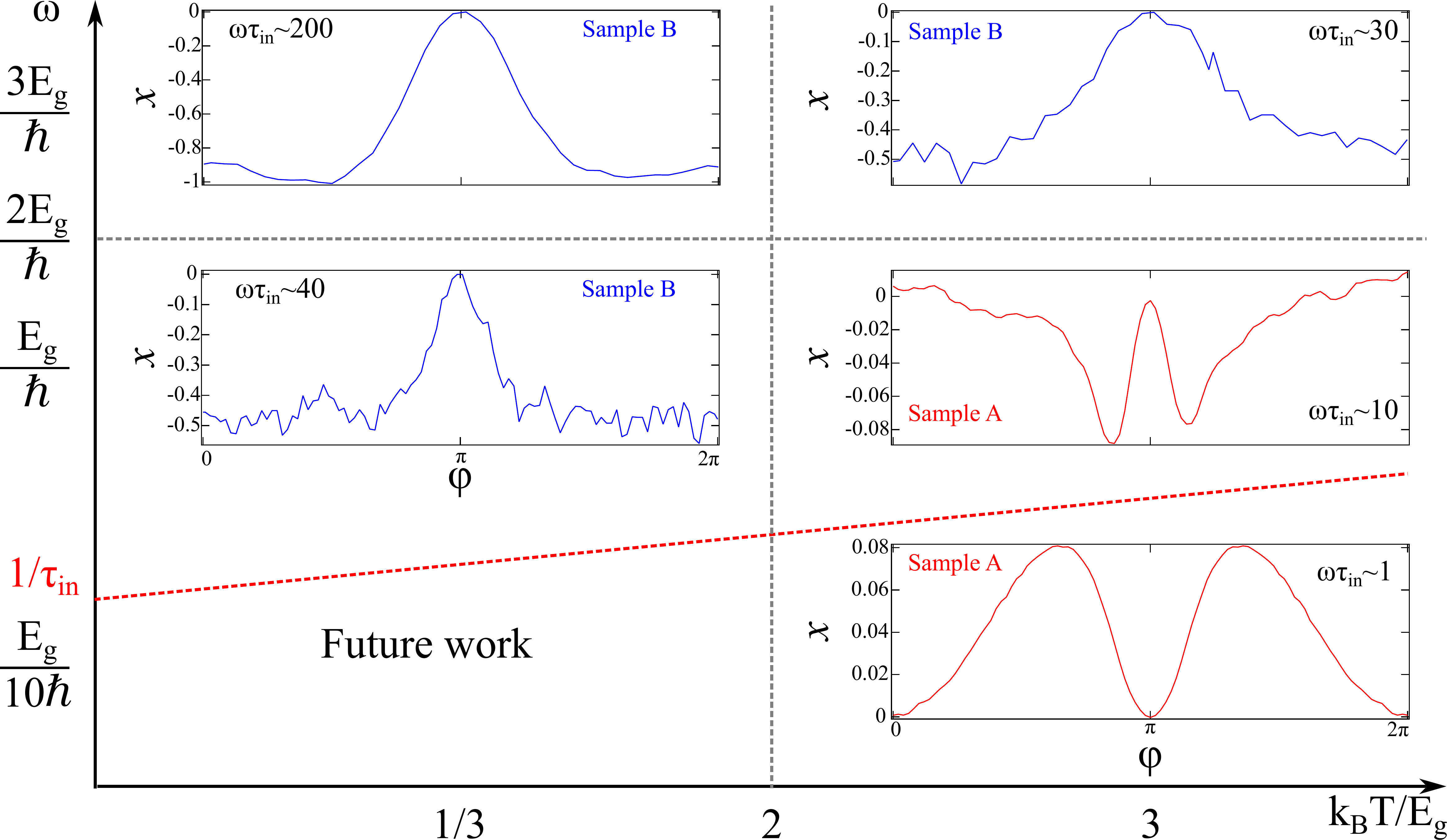}

\caption{\textbf{Summary of the characteristic phase dependence of $\chi''$ observed depending on the different energy scales in the system.} The normalized susceptibility $ x=\frac{\chi''(\varphi)-\chi''(\pi)}{2E_gG_N/\hbar} $ is plotted where $E_g/k_B=210~\rm{mK}$ and $ G_N=1~\rm{S} $ have been used for sample A and $ 2E_gG_N/\hbar=\delta_{\pi-0}\Xs(16\rm{GHz},\rm{80mK})  $ for sample B.}
\label{Figmeydi}
\end{figure}

To summarize, we have followed the frequency-dependent magnetic susceptibility of an NS ring  from the adiabatic regime, where the coherent response is simply the phase-derivative of the current-phase relation, to the highly non-adiabatic regime where dynamical effects lead to dissipation.  
When the frequency is close to the inelastic scattering rate, the relaxation of populations is the dominant dissipative process. This yields a phase-dependent dissipative response proportional to the inelastic time which is maximum around $ \pi/2$.  When the frequency is high enough   compared to the relaxation rate of the populations  $1/\tau_{in}$, dissipation is dominated by transitions across the minigap. Dissipation is then   related to the  dynamics of coherence (non-diagonal terms of the density matrix) but is independent of their relaxation time in the limit of a continuous spectrum. Due to the closing of the minigap, this process leads to an absorption peak at $\varphi=\pi$. In both cases dissipation measured by the real part of the admittance reaches values which are paradoxically much greater than the normal state conductance. We attribute this increase to enhanced transitions between electron-hole symmetric states. The phase dependences of these two contributions  contain different physics summarized in fig.\ref{Figmeydi}. Whereas the diagonal contribution is proportional to the square of the single level current, the non-diagonal contribution  is sensitive to  the phase-dependent minigap  at low temperature. At
high temperature compared to the minigap  the phase dependence  $\chi''(\varphi)$ is determined by the matrix elements of the current operator between different Andreev states.  These experiments  constitute therefore a powerful tool for the investigation
of the Andreev spectrum of a SNS junction. We propose to apply this method  to reveal  breaking of spin degeneracy, protected  crossings and topological superconductivity in Josephson  junctions with large spin-orbit interactions \cite{Glazman,vanHeck2017,Dmytruk2016,Murani2017,Trif2017}.

  We acknowledge A. Kasumov and F. Fortuna for help with the FIB and R. Danneau for suggesting the use of a conducting espacer. We are also grateful to M. Aprili, F. Chiodi, R. Deblock, M. Feigelman, T.T. Heikkil$ \ddot{\rm a} $, K. Tikhonov  and  P. Virtanen for fruitful discussions. 




\section{Appendix A}

\subsection{Overview of theoretical predictions.}
\label{sec:theo}
Following previous work on the dynamics of persistent currents in normal mesoscopic Aharonov-Bohm rings \cite{Trivedi1988,Reulet1994} we developed a Kubo formula description of the linear response of an NS ring to a time dependent Aharonov-Bohm flux.  In this approach, $ \chi $ splits naturally into three parts, $ \chi=\chi_J+\chi_D+\chi_{ND} $, as written in eq.\ref{eqchi} and detailed in the following.

\subsubsection{a)  Josephson susceptibility \mXj: adiabatic response }
In the adiabatic regime ($ \omega\rightarrow 0 $) the susceptibility  stems from the phase dependence of the ABS and does not depend on frequency. It is purely non-dissipative and is  the derivative of the Josephson supercurrent. At  temperature high enough to suppress its higher harmonics ($ k_{B}T\gtrsim E_g $)  the current-phase relation (CPR)
of  a long diffusive SNS junction is purely sinusoidal \cite{Dubos2001,Fuechsle2009} and the Josephson susceptibility reads 
\begin{equation}
 \chi_J=-\frac{2\pi}{\Phi_0}\frac{\partial I_J}{\partial \varphi}=-\frac{2\pi I_c}{\Phi_0}\cos(\varphi)
 \label{eqXj_HT}
\end{equation}
where $I_c(T)$  at high temperature \cite{DubosJLT} can be approximated by $I_c(T)= 
I_J(0) \exp(- k_BT/3.6 E_ {Th})$ where the zero temperature Josephson current reads: $I_J(0)= g_N 10.8 E_ {Th}/\Phi_0 $, with $g_N =G_Nh/2e^2$ is the dimensionless conductance.

In the following we discuss the extra contributions to the susceptibility when increasing frequency. Two characteristic frequency scales emerge the inverse relaxation time of the populations $\tau_{in} ^{-1}$ and  the inverse  diffusion time $\tau_{D} ^{-1}$, proportional to the minigap $E_g$ .
\subsubsection{b) Diagonal susceptibility $ \chi_D $: relaxation of ABS populations}
We discuss here $\chi_D$ called diagonal contribution because it only involves the diagonal elements of the current operator. It is the   non-adiabatic contribution  due to  the thermal relaxation of the populations  $f_n$ of the Andreev levels with the characteristic inelastic time $\tau_{in}$ . It is related to the phase-sensitivity of the spectrum and is therefore correlated to the existence of non-dissipative currents at equilibrium \cite{Buttiker1986}. Note that this contribution is specific to systems with a phase-dependent spectrum \cite{Trivedi1988} and is ignored in most derivations of the Kubo formula. To first approximation it  is proportional to the sum over an energy range $k_BT$   around the Fermi energy of the  square of the  single level current $i_n^2$.  
It can be recast as the product of a phase-independent term  by a frequency-independent  term 
\begin{equation}
\Xd(\omega,\varphi,T)=A(\omega,T)F(\varphi,T)
\label{eqXd_F}
\end{equation}
with  $A(\omega)=-\frac{i\omega }{\tin^{-1}-i\omega}$ and $\displaystyle  F(\varphi,T)= -\sum_n  i_n^2 \frac{\partial f_n}{\partial\epsilon_n } $. At high temperature compared to $E_{Th}$, $ F(\varphi,T) $ decays as $1/T$ due to the energy derivative of the Fermi distribution. $ F(\varphi,T) $ was first  introduced by Lempitsky \cite{Lempitskii1983} to describe the I(V) characteristics of  SNS junctions and was calculated numerically using time dependent  Usadel equations by Virtanen et al.\cite{Virtanen2011}. 

At high temperature compared to $E_{Th}$, $F(\varphi)$ can be approximated by the following analytical form: 
\begin{equation}
\begin{array}{l}
F_U(\varphi)\propto \frac{g_N}{k_BT} \frac{E_g^2}{\Phi_0^2}\\
\left[(-\pi + (\pi + \varphi)[2 \pi]) \sin (\varphi) - |\sin (\varphi)|\sin ^2(\varphi/2)/\pi \right].
\end{array} 
\end{equation}
It is dominated by its second harmonics  with in addition a sharp linear singularity at odd multiples of $\pi$. This is due to the dominant contribution  of Andreev levels close to the minigap whose flux dependence  is singular as in a highly transmitting superconducting  single channel point contact \cite{Martin-Rodero1996}.   It is  zero at $ \varphi=\pi $ and $ \varphi=0 $, 
 since single level currents go to zero at those  phases. It was recently shown that $\chi_D$ is very sensitive to the presence of  unavoided levels crossings in systems with spin-orbit interaction,  and can be  therefore useful to reveal topological superconductivity \cite{Murani2017}.  
\subsubsection{c) Non-diagonal susceptibility $ \chi_{ND} $:  microwave induced transitions }
We  now consider $\chi_{ND}$, the contribution that  describes microwave induced transitions within the Andreev spectrum. 
In the limit of a 	dense spectrum with level spacing $\delta_E \ll \hbar/\tau_{ND}$ , $\chi''_{ND} (\varphi)$ is independent of $\tau_{ND}$ and  reads:

\begin{equation}
\chi''_{ND}=-\omega \sum_{n,m \neq n} |J_{nm}|^2\frac{f_n  -f_m }{\epsilon_n-\epsilon_m} \delta(\epsilon_n-\epsilon_m-\hbar\omega ) 
\label{eqXsnd}
\end{equation}

that becomes in the continuous spectrum limit:
\begin{equation}
\begin{array}{l}
\chi''_{ND} =-\omega \displaystyle  \int_{}\int_{}d\epsilon d\epsilon' n(\epsilon, \varphi)n(\epsilon', \varphi) \\ \left[|J(\varphi,\epsilon,\epsilon' )|^2\displaystyle\frac{f(\epsilon) - f(\epsilon' )}{\epsilon-\epsilon'} 
\delta (\epsilon-\epsilon'-\hbar\omega)\right]
\end{array}{}
\label{eqXsNDap}
\end{equation}
where we have introduced  the phase-dependent density of states $n(\epsilon,\varphi)= n_0\left[ \theta(\epsilon-E_g(\varphi)) +\theta(-\epsilon-E_g(\varphi))\right]$ with $\theta$ the Heaviside distribution. $J(\varphi,\epsilon,\epsilon' )$ is the equivalent of $J_{nm}(\varphi)$ in the continuous limit.

We can distinguish two mechanisms involved in the phase dependence of $ \chi_{ND} $.  One is related to the phase dependence of the density of states whereas the second one is related to the phase dependence of the non-diagonal elements of the current operator $ J_{nm}(\varphi) $. When frequency is smaller or of the order of the minigap it is important to consider the phase dependence of the matrix elements  $ J_{nm}(\varphi) $  which  is determined by selection rules. In the case of a long and diffusive junction numerical simulations indicates that matrix elements between  electron-hole symmetric transitions are greater than  for non-symmetric ones \cite{Ferrier2013}; these selection rules are even more stringent in  ballistic systems or in the presence of spin-orbit interaction since no transitions are possible between Andreev states of opposite spin polarization \cite{vanHeck2017,Murani2017}.

\subsubsection{Low temperature}

Let's consider now the low temperature regime  such that $ E_g(\varphi) \gg k_B T $. No transitions are possible for $\hbar\omega < E_g(\varphi)$ and dissipation drops in the corresponding range of phase. 	 As a result in the limit of zero temperature and zero frequency dissipation  is expected to be a delta peak at $\pi$.  
This peak is found in the resolution of time-dependent Usadel equations \cite{Virtanen2011, Tikhonov2015} $ G(\pi) =  Y'(\pi) =\chi''(\pi)/\omega$   is found  increase at low temperature approximately like $G_N(1+A E_g/k_BT)$, with $ A $ a coefficient of the order of unity,  and to saturate  at temperatures such that $k_BT\lesssim \hbar\omega$.
Numerical simulations in \cite{Ferrier2013} show that the main contribution to this  dissipation peak at $\pi$ 
comes from  $J_{n,-n}$ non-diagonal matrix elements of the current operator connecting electron-hole symmetric states whose energy is of the order of the minigap.  We attribute the  increase of $ G(\pi,T) $ at low temperature  and low frequency to these enhanced transitions.

In contrast at high frequency, when $\hbar\omega \gg E_g(0) \gg k_BT$,  $\chi''_{ND}$ is dominated by the  high energy contribution of $J_{nm}$  which are  independent of phase and the phase-dependent absorption is exclusively determined by the density of states. It is therefore found to vary like:
\begin{equation} 
\chi''_{ND}(\varphi,\omega) = ( \omega -\frac{2  E_g(\varphi)}{\hbar})G_N 
\end{equation}
where we have used that  $<|J_{nm}|>^2 n^2(E_F) $   is the  classical Drude conductance $G_N=(2e^2/h)g_N $ of the normal wire. (The average $<|J_{nm}|>$ being taken on an energy scale much larger  than the superconducting gap but smaller than $\hbar/\tau_e$ where  $\tau_e$ is the elastic scattering time).
We find that at low temperature and high frequency compared to $E_g$  one   can reveal the minigap since  $\delta_{\varphi-0} \chi''_{ND}(\varphi,\omega) =-2G_N\frac{  E_g(\varphi)}{\hbar}$. It is independent of $\omega$ and $T$.

\subsubsection{High temperature}
In  the limit of  temperatures  higher than the minigap, the phase dependence of the density of states   can be neglected and  the phase dependence $\chi''_{ND}(\varphi)$ is  then dominated by the contributions  $\sum |J_{nm}(\varphi)|^2 \delta(\epsilon_n-\epsilon_m-\hbar\omega$). 
One can  then write:  $ \delta_\varphi \int d\omega \chi''_{ND}(\varphi,\omega)/\omega = \delta_\varphi\sum_{n,m \neq n} |J_{nm}(\varphi)|^2=\delta_\varphi\{ Tr(J^2)- \sum_n |J_{nn}(\varphi)|^2\} $



  Moreover, since  $Tr(J^2)$ is  phase-independent, the phase dependences of  $ \sum_{n \neq m}  |J_{nn}|^2 $ and  $ \sum_{n}|J_{nn}|^2 $ are  exactly opposite from one another.  
Numerical simulations \cite {Ferrier2013}  and time-dependent Usadel equations \cite{Tikhonov2015} indicate 	that  for $\hbar\omega\leq E_g \leq k_BT$ the phase dependence of $\chi''_{ND}$  is independent of frequency and  varies like $\delta_{\varphi-0}\chi''_{ND} = (-\hbar\omega/E_g) F_U(\varphi) $. 

\subsubsection{d) Relations between dissipative and non-dissipative  phase-dependent contributions}
In the following we show that simple relations hold between the dissipative and non-dissipative contributions of $\delta _{\pi-0}\chi$ when $\omega \gg \tau_{in}$. They can be obtained from eq.\ref{eqXsNDap} for $\chi''_{ND}$ and the following expression of $\chi' $:  

\begin{equation}
\begin{array}{l}
\delta_{\varphi-0} \chi' =-\delta_{\varphi-0} \displaystyle\int \int_{}d\epsilon d\epsilon' n(\epsilon, \varphi)n(\epsilon', \varphi)\\ \left[|J(\varphi,\epsilon,\epsilon' )|^2\displaystyle\frac{f(\epsilon) - f(\epsilon' )}{\epsilon-\epsilon'}\right] 
\end{array}{}
\label{eqNDap}
\end{equation}


The phase-dependent contribution  $\delta_{\varphi-0}|J (\varphi,\epsilon,\epsilon' )|^2 n(\epsilon, \varphi)n(\epsilon', \varphi)$ is approximated by $g^+(\epsilon^+/\epsilon_c^+)\times g^-(\epsilon^-/\epsilon_c^-)$  where $g^+$ and $g^-$   are functions of $\epsilon^+=(\epsilon+\epsilon')/2$ and  $\epsilon^-=(\epsilon-\epsilon')/2$.
The  correlation functions $g^+$ and $g^-$ are expected to  oscillate and decay on energy 
scales  $\epsilon^+ _c$ and $\epsilon^-_c$  of the order of $E_g$. 
Moreover  in the limit where $k_BT \gg \epsilon^-$ one can  approximate in both integrals  the function $\frac{f(\epsilon) - f(\epsilon' )} {\epsilon^-}$  by a square function centered on  $\epsilon^+$ of width $4k_BT$ which yields for $\hbar\omega \ll E_g \ll k_BT$:

\begin{equation}
\begin{array}{l}

\delta_{\varphi-0}\chi''(\omega) \simeq \omega  G_N  \int_0 ^{2k_BT}   g ^+(\epsilon^+/\epsilon_c^+) d\epsilon^+\\
\delta_{\varphi-0}\chi' \simeq G_N/\hbar\int_0 ^\infty g^- (\epsilon^-/\epsilon_c^-)  d\epsilon^- \int_0 ^{2k_BT}   g ^+(\epsilon^+/\epsilon_c^+) d\epsilon^+ 

\end{array}{}
\label{eqND2}
\end{equation}

From these expressions it is possible to deduce a simple general  relation between $
\delta_{\varphi-0}\chi''_{ND}$ and $\delta_{\varphi-0}\chi'= \delta_{\varphi-0}\chi_J$ valid for $\omega\tau_{in}\gg 1$ .

\begin{equation}  
\delta_{\varphi-0}\chi''_{ND}=(\hbar\omega/\epsilon_c^-)\delta_{\varphi-0}\chi'\simeq (\hbar\omega/E_g)\delta_{\varphi-0}\chi'(T,\varphi)
\end{equation} 
In particular for $\varphi=\pi$  we can use that $\delta_{\pi-0}\chi'_D =0$ and  $\delta_{\pi-0}\chi' =\delta_{\pi-0}\chi_J\propto \exp (-k_BT/1.16 E_g)$. $\delta _{\pi-0}\chi''_{ND}$ is therefore   expected to decrease exponentially with T like the Josephson current. For other values of $\varphi$, assuming that $\epsilon_c^- \simeq E_g$ this result yields the expected phase dependence of $\delta \chi''_{ND}(\varphi) =  -(\omega/ E_g)  \delta\chi'_D (\varphi) =-(\omega/ E_g ) F_U(\varphi)$ in this range of parameters, $\hbar\omega \ll E_g \ll k_BT$, leading to a maximum of $\delta \chi''_{ND}(\varphi)$ of the order of:
\begin{equation}
\delta \chi''_{ND} =(I_J(0)/\Phi_0) (\hbar\omega /k_BT) = G_N \omega E_g/(k_BT)
\end{equation}
i.e. much higher than $\delta _{\pi-0}\chi''_{ND}$. 

 These relations  between $\delta_{\varphi-0}\chi''_{ND}$ and $\delta_{\varphi-0}\chi'$ are useful for the interpretation of our experimental data in the range of parameters where screening effects distort the phase dependence of $\chi''_m$. They can be understood as the Kramers-Kronig relation between the dissipative and non-dissipative components of the phase-dependent susceptibility. Extrapolating this reasoning to lower temperature and higher frequency, we expect that the ratio $\delta_{\pi-0}\chi''_{ND}(\omega)/\delta_{\pi-0}\chi'$  will increase at low temperature  (in agreement with the experimental results on sample B, see fig.\ref{fig:evo_X_T_f}) but  its  calculation is complicated by logarithmic singularities in $\chi'$,  and is beyond the scope of this paper.
 
 \section{Appendix B: Determination of the absolute value of $\chi''(\pi)$}
 
  We focus on the range of parameters such that $L_l\chi''(\varphi)$ can be neglected in eq.\ref{chis} leading to simplified relations between $\chi_m$ and $\chi$:
	
\begin{equation}
\begin{array}{l}
\chi'_m(\varphi)= \chi'(\varphi)/(1 - L_l\chi'(\varphi))
\\
\chi''_m (\varphi) =\chi''(\varphi)/(1-L_l\chi'(\varphi))^2
\end{array}{}
\end{equation}


  From these relations one can easily deduce the ratio $R_m=  \delta_{\pi-0} \chi''_m/ \delta_{\pi-0}\chi'_m $. This quantity  is interesting since it does not  depend on the coupling  coefficient $L_c^2$  between the  measured and intrinsic  susceptibility which is not easy to estimate accurately. It reads: 

\begin{equation}
R_m=\frac{2 \chi''(\pi,T) L_l}{1-\beta^2}+\frac{ \delta_{\pi-0}\chi''}{\delta_{\pi-0}\chi'}\frac{1-\beta}{1+\beta}
\label{eqRm}
\end{equation}

The first contribution to $R_m$  is proportional to $L_l$   and thus is due to screening.  The  denominator  contains a term in $\beta ^2$ which is negligible in the range of temperature we have investigated ($\beta\leq 0.4)$. The second contribution  is proportional to the intrinsic ratio between $\delta_{\pi-0}\chi''$ and $\delta_{\pi-0}\chi'$. Theoretical predictions (see Appendix A) indicate that it is reasonable to consider that $ \chi''(0)=\chi''(\pi)$ when $ T>2 E_g $, we therefore assume $\delta_{\pi-0}\chi''=0$. Under this assumption and within first order in $\beta$ equation \ref{eqRm} yields:
\begin{equation}
R_m=2 \chi''(\pi,T) L_l
\label{eqRm2}
\end{equation}

Therefore we have shown how to deduce   the value of $\chi''(\pi,T)= R_m (T)/2 L_l$ shown in fig.\ref{fig:Xss} from the measurement of $R_m (T)$ and the estimation of $L_l$.  



 \section{Appendix C: high frequency data on sample A, for which screening corrections cannot be quantitatively  determined}

\begin{figure}[h!]
	\centering
	\includegraphics[width=\linewidth]{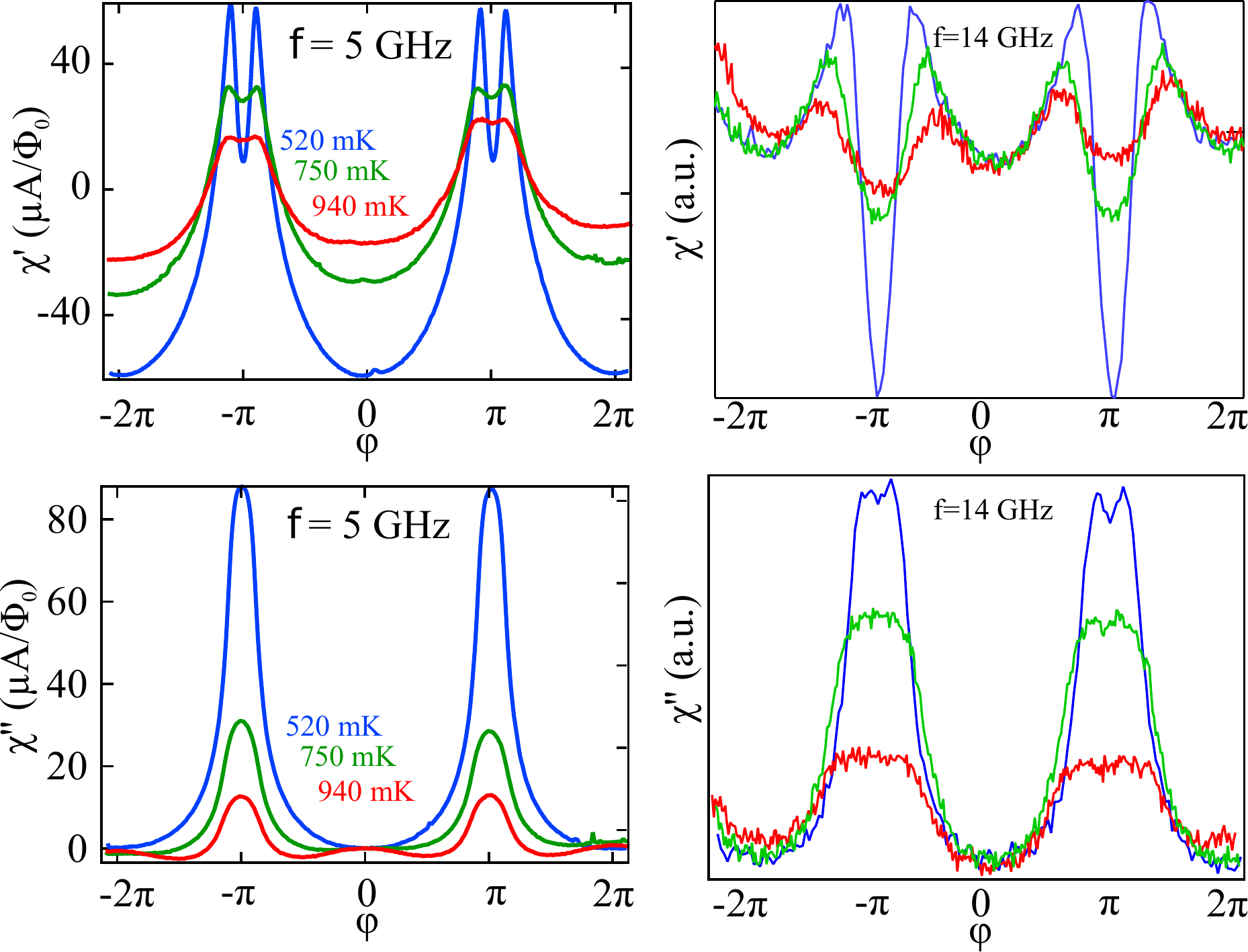}
	\caption{Evolution of the  phase dependence of $ \chi '_m $ (top) and $ \chi"_m $ (bottom) with temperature at 5 GHz and 14 GHz measured on sample A. Dips in $ \chi'_m $ at $ \pi $ are attributed to screening effects. Data in this regime could not be corrected.}
	\label{fig:mix_exp}
\end{figure}

\begin{figure}[h!]
	\centering
	\includegraphics[height=2.8cm]{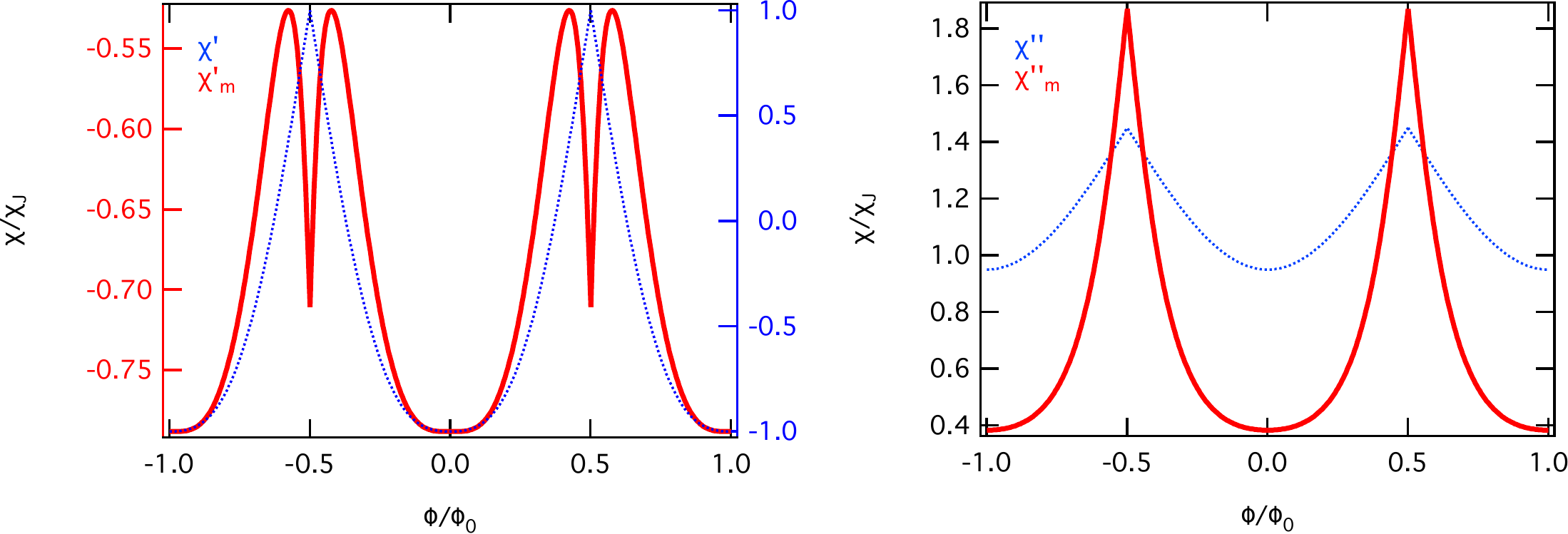}
	\caption{ Computed phase dependences of $\chi'_m$ and $\chi''_m$ assuming $\chi'=-\cos \varphi + \chi_{D}$ and $\chi''(\varphi)$ similar to $E_g(\varphi)$. One sees that $\chi'_m$ exhibits a dip at $\pi$ and  what $\chi''_m$ is much larger than $\chi''$ when  mixing between $\chi'$ and $\chi''$  contributions  in $\chi'_m$ and $\chi''_m$ (see Eqs.\ref{eq:df_Xp} and \ref{eq:dQ_Xs}) becomes important.}
	\label{fig:mix_theo}
\end{figure}

For very high frequency (above 5 GHz) the screening corrections quadratic  in $ L_l\chi''$ in equations \ref{chim} and \ref{chis} cannot be neglected   on sample A  and give rise to an important intrication  between the phase dependence of  $\chi'(\varphi)$ and $\chi''(\varphi)$ at low temperature and high frequency. This is illustrated below  where  $\chi'_m$ exhibits  a sharp split peak at $\pi$ as seen in fig.\ref{fig:mix_exp}.    This intriguing phase dependence can be shown to be due to the negative term in expression \ref{eq:dQ_Xs} proportional to $ -\chi''(\varphi)^2$ as illustrated in fig. \ref {fig:mix_theo}. Correlatively $\chi''_m(\varphi)$ exhibits a very sharp increase at $\pi$
 
\end{document}